\documentclass[aps,pra,showpacs,twocolumn,superscriptaddress]{revtex4}
\usepackage{amssymb}
\usepackage{amsmath}
\usepackage{graphicx}
\usepackage{txfonts}

\begin{document}

\title{Steady-state phase diagram of a driven QED-cavity array with cross-Kerr nonlinearities}

\author{Jiasen Jin}
\affiliation{NEST, Scuola Normale Superiore and Istituto di Nanoscienze-CNR, I-56126 Pisa, Italy}

\author{Davide Rossini}
\affiliation{NEST, Scuola Normale Superiore and Istituto di Nanoscienze-CNR, I-56126 Pisa, Italy}

\author{Martin Leib}
\affiliation{Technische Universit\"at M\"unchen, Physik Department, James-Franck-Str., D-85748 Garching, Germany}

\author{Michael J. Hartmann}
\affiliation{Institute of Photonics and Quantum Sciences, Heriot-Watt University, Edinburgh, EH14 4AS, United Kingdom}

\author{Rosario Fazio}
\affiliation{NEST, Scuola Normale Superiore and Istituto di Nanoscienze-CNR, I-56126 Pisa, Italy}
\affiliation{Centre for Quantum Technologies, National University of Singapore, 3 Science Drive 2, Singapore 117543}

\begin{abstract}
We study the properties of an array of QED-cavities coupled by nonlinear elements
in the presence of photon leakage and driven by a coherent source. The main effect of
the nonlinear couplings is to provide an effective cross-Kerr interaction between
nearest-neighbor cavities. Additionally correlated photon hopping between neighboring
cavities arises. We provide a detailed mean-field analysis of the steady-state phase diagram
as a function of the system parameters, the leakage and the external driving,
and show the emergence of a number of different quantum phases. A {\it photon crystal}
associated to a spatial modulation of the photon blockade appears. The steady state can also
display oscillating behavior and bi-stability. In some regions the crystalline ordering may
coexist with the oscillating behavior. Furthermore we study the effect of short-range quantum
fluctuations by employing a cluster mean-field analysis. Focusing on the corrections to the
photon crystal boundaries, we show that, apart for some quantitative differences, the cluster
mean field supports the findings of the simple single-site analysis. 
In the last part of the paper we concentrate on the possibility to build up the class 
of arrays introduced here, by means of superconducting circuits of existing technology. 
We consider a realistic choice of the parameters for this specific implementation 
and discuss some properties of the steady-state phase diagram.
\end{abstract}


\maketitle

\section{Introduction}

Since its beginning, the study of light-matter interaction in cavity and circuit quantum electrodynamics (QED)
has been providing a very fertile playground to test fundamental questions at the heart
of quantum mechanics, together with the realization of very promising implementations
of quantum processors~\cite{raimond2001,girvin2009}.
Recently the attention has also focused on the study of systems where photon hopping
between neighboring cavities introduces an additional degree of freedom
and leads to a wealth of new phenomena.
The topic has been already reviewed in a number of works~\cite{hartmann2008,tomadin2010,houck2012,carusotto2013,schmidt2013}
and the first experimental results on cavity arrays are beginning to appear~\cite{underwood2013,Abbarchi13,Toyoda13}.
Very recently a dissipation-driven phase transition in a coupled cavity dimer has been
reported~\cite{raftery2013} which paves to similar studies on larger arrays.

Cavity arrays are periodic arrangements of QED-cavities aimed at studying many-body states with photons.
In their first conception~\cite{hartmann2006,greentree2006,angelakis2007},
as well as in most of the subsequent papers on the topic, the coupling between
neighboring cavities has been mediated by photon hopping.
Indeed it was envisaged, and confirmed by an extensive number of works,
that a very rich phenomenology arises from the interplay between hopping and
strong local nonlinearities leading to photon blockade~\cite{imamoglu97,rebic99,kim99,birnbaum05}.

As long as particle losses can be ignored, the properties of cavity arrays resemble
in several aspects those of the Bose-Hubbard model~\cite{fisher1989}.
In the photon-blockade regime the cavity array enters a Mott phase, in which photon
number fluctuations are suppressed. On the contrary, when the hopping between neighboring
cavities dominates over the local nonlinearities, photons are delocalized through the whole array.
In the absence of leakage, photon number is conserved and this phase has long-range
superfluid correlations. Some care with the definition of superfluidity in photonic systems
has to be taken in the (more realistic) case where leakage is present.
A discussion of this issue has just been started for the case of cavity arrays~\cite{RuizRivas14},
see also Refs.~\cite{keeling2011,wouters2010} for an analysis in related systems.
The ``equilibrium'' phase diagram of coupled cavity arrays has been thoroughly studied and the location
of the different phases, together with the critical properties of the associated
phase transitions, have been determined (A fairly complete review of our present
understanding can be found in Refs.~\cite{hartmann2008,tomadin2010,houck2012,schmidt2013}.).

Cavity arrays, however, naturally operate under nonequilibrium conditions,
{\it i.e.}, subject to unavoidable leakage of photons which are pumped back
into the system by an external drive. In that case, the situation may change drastically,
and it is to a large extent an unexplored territory.
Only very recently the many-body nonequilibrium dynamics of cavity arrays started to be
addressed---see, {\it e.g.}, 
Refs.~\cite{carusotto2009,tomadin2010b,Hartmann10,Nunnenkamp2011,nissen2012,grujic2012,grujic2013,jin2013,RuizRivas14,leboite2013,yuge2013} 
and references therein---thus entering the exciting field of quantum phases and phase transitions
in driven quantum open
systems~\cite{diehl2008,diehl2010,baumann2010,lesanovsky2010,lee2011,ludwig2012,tomadin2012}.
In this paper we further pursue this direction and study the steady-state properties
of a cavity array in the presence of photon leakage and subject to an external
uniform coherent drive. The additional new ingredient we introduce
is a cavity coupling through nonlinear elements.

So far, with some notable exceptions~\cite{zueco2012, zueco2013, jin2013},
the coupling between cavities has been considered only through photon hopping.
Implementations based on circuit-QED~\cite{wallraff2004} however provide enough flexibility
to connect two neighboring cavities both via linear ({\it e.g.}~capacitors)
and nonlinear ({\it e.g.}~Josephson nano-circuits) elements.
This freedom paves the way to explore
a multitude of different engineered Hamiltonians with systems of cavities.
At this point, it is also worth to stress that the implementation of cavity arrays
within circuit-QED is very promising, the first experiments with arrays
of up to five cavities have been done~\cite{lucero2012,Steffen13,Chen14}, 
and experiments with lattices of cavities are progressing fast~\cite{houck2012}.

In the present paper we expand on the results discussed in Ref.~\cite{jin2013}.
We will give a more detailed account on the steady-state phase diagram.
Most of the analysis will be performed by means of a single-site mean-field decoupling.
We will further check the robustness of our results by performing a cluster mean-field analysis
to take into account the effect of short-range quantum fluctuations.

The paper is organized as follows.
In the next Section~\ref{model} we define the model and its dynamics, dictated by both
the unitary evolution (which includes the external drive) and the dissipation.
In Section~\ref{meanfield} we introduce the mean-field approximation which will then be used
to extract the steady-state phase diagram, extensively described in Section~\ref{diagram}.
In Section~\ref{clusterm} we include the effect of short-range correlations by performing
a cluster mean-field analysis. We conclude by discussing in Section~\ref{cQED} some specific
aspects of the implementation with circuit-QED cavities. A summary of our results is given in
Section~\ref{conclusions}.

\section{The model}
\label{model}

The model we will investigate, including an external coherent drive, 
is described by the Hamiltonian~\cite{jin2013}
\begin{eqnarray}
  {\cal H} & = & \sum_{i}{\left[-\delta n_i + \Omega(a_i+a_i^\dagger)\right]} - J\sum_{\langle i,j \rangle}{(a_i^\dagger a_j + \text{H.c.})}\nonumber \\
  && + U\sum_{i}{n_i(n_i-1)} + V \sum_{\langle i,j \rangle}{n_in_{j}}\nonumber \\
  && + \sum_{\langle i,j \rangle}{\left[\frac{J_2}{2} a_i^\dagger a_i^\dagger a_j a_j
      - J_n a_i^\dagger (n_i+n_j) a_j + \text{H.c.} \right]} \; ,
  \label{modelham}
\end{eqnarray}
in the rotating frame with respect to the frequency of the external drive.
The number operator $n_i=a^\dagger_ia_i$ counts the photons in the $i$-th cavity,
$\delta$ is the detuning of the cavity mode with respect to the frequency of the pump,
and $\Omega$ is the amplitude of the coherent pumping. The term proportional to $J$ is the standard rate
for the hopping of individual photons between neighboring cavities.
The two contributions in the second line take into account the effective (Kerr) interaction
between the photons: $U$ quantifies the onsite repulsion, while $V$ is the cross-Kerr nonlinearity.
The remaining terms describe correlated photon hopping.
The term proportional to $J_2$ is responsible for the pair hopping
and the term proportional to $J_n$ describes the hopping to a neighboring cavity
controlled by the occupation of that cavity.
The brackets $\langle \cdot, \cdot \rangle$ indicate that the sum is restricted to nearest neighbors.

In addition to the unitary part, there are losses due to photons leaking out of the cavities.
The dynamics of the density matrix $\rho$ of the system is then governed by the master equation
\begin{equation}
  \dot{\rho} = -i[{\cal H},\rho]+\frac{\kappa}{2}\sum_i{(2a_i\rho a_i^\dagger-n_i\rho-\rho n_i)} \,,
  \label{mastereq}
\end{equation}
where $\kappa^{-1}$ is the photon lifetime. Hereafter we set $\kappa=1$,
and work in units of $\hbar = 1$.

\section{Single-site mean-field decoupling}
\label{meanfield}

Solving exactly the dynamics dictated by Eq.~\eqref{mastereq} is a formidable task.
Here we study the steady-state phase diagram by employing a mean-field decoupling,
which should become accurate in the limit of arrays with large coordination number $z$.
There are several terms in the Hamiltonian that involve couplings between different sites:
besides the hopping, including single-, two-photon and correlated hopping, one also has to take
into account the cross-Kerr term.
Indeed the latter contribution (controlled by $V$) may favour the stabilization of
a {\it photon crystal} phase, in which the photon blockade is density modulated~\cite{jin2013}.
In order to include this possibility, the decoupling
should be different for different sublattices.
Let us consider a bipartite lattice:
indicating with $A$ and $B$ the corresponding sublattices,
the different terms are decoupled as follows:
\begin{eqnarray}
  z^{-1}\sum_{\langle i,j \rangle}{n_in_j} & \longrightarrow & \langle n_A\rangle \sum_{j\in B}{n_j} + \langle n_B\rangle\sum_{i\in A}{n_i},\cr\cr
  z^{-1}\sum_{\langle i,j \rangle} {a_i^\dagger a_j} & \longrightarrow & \langle a_A^\dagger\rangle\sum_{j\in B}{a_j}+
  \langle a_B^\dagger\rangle\sum_{i\in A}{a_i},\cr\cr
  z^{-1}\sum_{\langle i,j \rangle}{a_i^\dagger a_i^\dagger a_j a_j} & \longrightarrow & \langle a_A^\dagger a_A^\dagger\rangle\sum_{j\in B}{a_j a_j}
  +\langle a_B^\dagger a_B^\dagger\rangle\sum_{i\in A}{a_i a_i},\cr\cr
  z^{-1}\sum_{\langle i,j \rangle}{a_i^\dagger n_ia_j} & \longrightarrow & \langle a_A^\dagger n_A\rangle\sum_{j\in B}{a_j}+\langle a_B^\dagger
  n_B\rangle\sum_{i\in A}{a_i},\cr\cr
  z^{-1}\sum_{\langle i,j \rangle}{a_i^\dagger n_j a_j} & \longrightarrow & \langle a_A^\dagger\rangle\sum_{j \in B}{n_j a_j}+
  \langle a_B^\dagger\rangle\sum_{i\in A}{n_i a_i}. \label{Eq_meanfield}
\end{eqnarray}
After the decoupling, the density matrix of the array can be written as
$\rho = \prod_{i \in A} \rho_i \prod_{j \in B} \rho_j$.
Moreover in each sublattice the system can be considered as uniform.
The dynamics is thus reduced to two coupled equations for the density matrices
of the $A$ and $B$ sublattices, respectively $\rho_A$ and $\rho_B$,
\begin{eqnarray}
  \dot{\rho}_A&=&-i[\mathcal{H}_A,\rho_A]+\frac{\kappa}{2}(2a_A\rho_A a_A^\dagger-n_A\rho_A-\rho_A n_A),\label{twositesmaster1}\\
  \dot{\rho}_B&=&-i[\mathcal{H}_B,\rho_B]+\frac{\kappa}{2}(2a_B\rho_B a_B^\dagger-n_B\rho_B-\rho_B n_B),\label{twositesmaster2}
\end{eqnarray}
with
\begin{eqnarray}
  \mathcal{H}_A & = & -\delta n_A + \Omega(a_A + a^\dagger_A) + Un_A(n_A-1) + z V w_B n_A \nonumber \\
  &&-zJ(\psi_Ba^\dagger_A+ \text{H.c.}) +\frac{zJ_2}{2}(\phi_Ba^\dagger_Aa^\dagger_A+\text{H.c.}) \nonumber \\
  &&-zJ_n(\chi_B a^\dagger_A+\text{H.c.})-zJ_n(\psi_Ba^\dagger_An_A+\text{H.c.}), \\
  \mathcal{H}_B & = & -\delta n_B + \Omega(a_B + a^\dagger_B) + Un_B(n_B-1) + z V w_A n_B \nonumber \\
  &&-zJ(\psi_Aa^\dagger_B+\text{H.c.}) + \frac{zJ_2}{2}(\phi_Aa^\dagger_Ba^\dagger_B+\text{H.c.}) \nonumber \\
  &&-zJ_n(\chi_A a^\dagger_B+ \text{H.c.})-zJ_n(\psi_Aa^\dagger_Bn_B+\text{H.c.})
  \label{twositesHamiltonian}
\end{eqnarray}
where $w_i=\mathrm{tr}(n_i \rho_i)$, $\psi_i=\mathrm{tr}(a_i \rho_i)$, $\phi_i=\mathrm{tr}(a_i a_i \rho_i)$, and
$\chi_i=\mathrm{tr}(n_i a_i \rho_i)$ for $i=A,B$.

A difference in the average photon population of the two sublattices,
$\langle n_A \rangle \ne \langle n_B \rangle$, signals a crystalline phase
in which the $A$-$B$ symmetry is spontaneously broken.
This quantity can be used as an order parameter.
On the contrary, an average $\langle a_A \rangle$ or $\langle a_A a_A \rangle$
(or equivalently $\langle a_B \rangle$ or $\langle a_B a_B \rangle$)
different from zero cannot be associated to any superfluid ordering in the steady state.
In this sense the situation is very different from the ``equilibrium''
scenario (neither damping nor driving).
Due to the external coherent driving, there is no spontaneous breaking
of the $U(1)$ gauge symmetry. The external drive induces a global coherence,
that is not at all related to the collective behavior of the cavity array.
For an incoherent pumping, in contrast, spontaneous coherence can be found
but does not lead to a non-vanishing expectation value $\langle a \rangle$ for the field
(see Ref.~\cite{RuizRivas14}).

Related to this last point, it is worth commenting on the correlated hopping
appearing in the last line of Eq.~\eqref{modelham}: the corresponding decoupling is given in the last three lines
of the mean-field approximation, Eq.~\eqref{Eq_meanfield}.
In Ref.~\cite{sowinski2012} it was shown that the presence of these contributions
in the Hamiltonian may considerably enrich the phase diagram.
In the present context however they will not play a major role.
In the physical implementation we consider, the associated coupling constants
$J_2$ and $J_n$ are parametrically smaller then the single-photon hopping. We will mostly analyze regions of the
phase diagram where $J_2$, $J_n$ $\ll J$.
Therefore, as we will see, they will lead only to some quantitative modifications
of the phase boundaries. This is not however a fundamental issue:
In circuit-QED implementations for example, it is possible to fine tune the values of the circuit elements
in order to suppress the single-photon hopping in favor of the correlated one.
Concerning the comparison with the cross-Kerr term, it is not difficult to imagine
Josephson circuits that will lead to an enhancement of the correlated hopping.
There is however a second reason why the results of Ref.~\cite{sowinski2012} cannot be
directly applied here. It is again related to the fact that under nonequilibrium conditions,
as we analyze here, coherence is built up because of the external drive.
It would be very interesting to explore in this context situations where superfluidity
in cavity arrays arises spontaneously~\cite{RuizRivas14}. In such circumstances, nonlinear couplings
that introduce correlated hoppings could lead to very interesting photon-pair superfluidity.

In the next Section we discuss the properties of the steady-state phase diagram by solving
Eqs.~\eqref{twositesmaster1}-\eqref{twositesmaster2} for various choices of the couplings.

\section{Phase diagram}
\label{diagram}

The phase diagram derived from Eqs.~\eqref{twositesmaster1}-\eqref{twositesmaster2} is quite rich.
A first account of this was given in Ref.~\cite{jin2013}.
Here we extend the analysis and provide some more detailed discussions.
Due to the cross-Kerr nonlinearity, the steady-state phases
can be classified into uniform and checkerboard phases.
This is a general feature of the exact model and it is captured in the mean-field approximation
by introducing a decoupling which takes into account different averages on different sublattices.
As already mentioned, these two phases can be distinguished by the order parameter
$\Delta n=|\langle n_A\rangle-\langle n_B\rangle|$. Here we refer only to the steady state,
therefore the value of $\Delta n$ is time-independent, unless stated otherwise (see below).
In the uniform phase, the steady-state photon population in the two sublattices
is identical, which means a vanishing $\Delta n$.
In checkerboard (crystalline) phase the photon number in the cavity array is modulated
as in a photon crystal. The photon population of one sublattice is higher
than that of the other one, which means a non-zero $\Delta n$. Yet, this is not the whole story.

For some values of the coupling constants, the observables can never
be time-independent even in the long-time limit.
Instead, the system will enter an oscillatory phase in which the photon number
of each sublattice oscillates periodically with $\langle n_A\rangle\ne\langle n_B\rangle$.
The steady state can further show bistable behavior and dependence of the initial conditions.
All of this can occur both in the uniform and in the crystalline phases.
The richness of the steady-state phase diagram arises due to all these combinations which can appear.
In order to simplify the presentation, the discussion has been organized
in different sections, for various classes of values of the couplings.
In the following we choose different values of the couplings as compared to~\cite{jin2013}.
When not specified, the correlated hopping terms are set to zero.

\subsection{Infinite onsite interaction ($U \to \infty$)}

In the limit $U= +\infty$, $J_2= 0$ and $J_n = 0$, we recover the model studied
by Lee {\it et al.}~\cite{lee2011}.
In fact, when $U$ represents the largest energy scale in the problem,
our phase diagram coincides with that of Ref.~\cite{lee2011}.
Note however that, differently from what is usually encountered in other systems
with extended Hubbard-like interaction, for circuit-QED implementations as discussed in Section~\ref{cQED}, 
the case in which $U \le V$ makes sense as well.
As long as the onsite repulsion $U$ is much larger than the other
energy scales (except possibly of $V$) it is always possible to reduce
the local Hilbert space to only two states and the results of Ref.~\cite{lee2011} apply.

\subsection{Zero onsite interaction ($U = 0$)}

The situation in which both the onsite interaction and the correlated hopping vanish ($U = J_2 = J_n=0$)
can be solved exactly, within the mean-field approximation.
The coupled master equations in Eqs.~\eqref{twositesmaster1}-\eqref{twositesmaster2}
can be rewritten in the form of complex differential equations as follows,
\begin{eqnarray}
  \dot{w}_A & = & 2\Omega y_A + 2 z J x_A y_B - 2 z J y_A x_B - w_A, \nonumber \\
  \dot{x}_A & = & -(-\delta + z V w_B) y_A + z J y_B - x_A/2,\nonumber \\
  \dot{y}_A & = & (-\delta + z V w_B) x_A - z J x_B + \Omega - y_A/2 \nonumber \\
  \dot{w}_B & = & 2\Omega y_B + 2 z J x_B y_A - 2 z J y_B x_A - w_B, \nonumber \\
  \dot{x}_B & = & -(-\delta + z V w_A) y_B + z J y_A - x_B/2, \nonumber \\
  \dot{y}_B & = & (-\delta + z V w_A) x_B - z J x_A + \Omega - y_B/2,
\label{dfeq}
\end{eqnarray}
where $x_{j}$ and $y_{j}$ are the real and imaginary parts
of $\psi_{j}^*$ [{\it i.e.}~$\text{Tr}( a_j \rho_j) = x_{j} - i \, y_{j}, j=A,B$], respectively.
We focus on the fixed points of the system, {\it i.e.}, when $\dot{w}_{A,B}=\dot{x}_{A,B}=\dot{y}_{A,B}=0$.

\subsubsection{Zero hopping}

When $J=0$, the non-uniform fixed points are given by
\begin{equation}
\begin{array}{lll}
    w_{A} & \displaystyle = \frac{2p_{A}}{zV},
    & \displaystyle w_{B} = \frac{2p_{B}}{zV}, \vspace{1mm} \\
    x_{A} & \displaystyle = \frac{8\delta p_{A}-4\delta^2-1}{4zV\Omega},
    & \displaystyle x_{B} = \frac{8\delta p_{B}-4\delta^2-1}{4zV\Omega},\vspace{1mm} \\
    y_{A} & \displaystyle = \frac{p_{A}}{zV\Omega},
    & \displaystyle y_{B} = \frac{p_{B}}{zV\Omega},
  \end{array}
\end{equation}
where $p_A$ and $p_B$ are two different real roots of the quadratic equation
\begin{equation}
  16 \gamma p^2 - 16 (\gamma \delta + 2 z V \Omega^2) p + \gamma^2 = 0 \,,
  \label{quadraeq}
\end{equation}
with $\gamma = 4 \delta^2 + 1$.
The non-uniform fixed points exist only when $V$ satisfies the condition,
\begin{equation}
  zV > \frac{\gamma ( \sqrt{\gamma} - 2 \delta)}{4 \Omega^2} \,.
\end{equation}
On the other side, the uniform fixed points are given by
\begin{equation}
  \begin{array}{lllll}
    w_A & = & w_B & = & 2\Omega \bar{p}, \vspace{0.5mm} \\
    x_A & = & x_B & = & 2\bar{p}(\delta-2zV\Omega \bar{p}),\vspace{0.5mm} \\
    y_A & = & y_B & = & \bar{p},
  \end{array}
\end{equation}
where $\bar{p}$ is any possible positive real root of the equation
\begin{equation}
  16 z V \Omega p^2 (z V \Omega p - \delta) + \gamma p - 2 \Omega = 0 \,.
  \label{cubiceq}
\end{equation}
Since Eq.~\eqref{cubiceq} is a cubic equation in $p$,
the number of positive real roots can be determined by Descartes' rule of signs.
We see that for $\delta\le 0$ there is only one uniform fixed point
and for $\delta>0$ there might be one or three uniform fixed points.
Furthermore, if there exist three uniform fixed points, the polynomial corresponding
to Eq.~\eqref{cubiceq} should have a positive local maximum and a negative local minimum.
Thus there are three uniform fixed points if and only if
\begin{equation}
  \delta > \frac{\sqrt{3}}{2} \,, \quad \mbox{and} \quad
  \bigg\vert zV - \frac{\delta \xi + 12 \delta}{54 \, \Omega^2} \bigg\vert < \frac{\xi^{3/2}}{108 \, \Omega^2} \,,
\end{equation}
where $\xi=4\delta^2-3$.

\begin{figure}[t]
  \includegraphics[width = 1\linewidth]{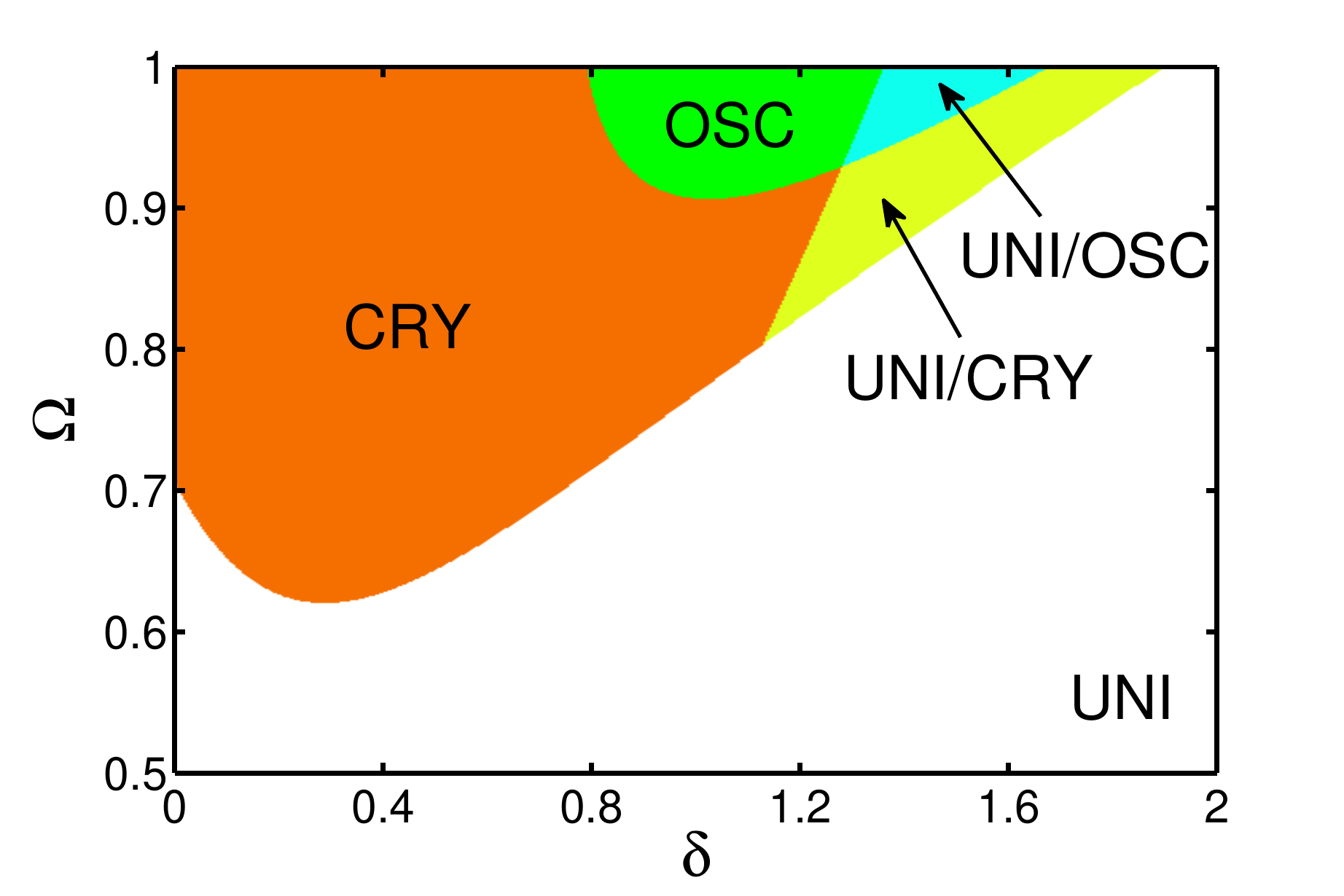}
  \caption{(Color online). Phase diagram of the system with $U=J=J_2=J_n=0$, and $zV=0.5$.
    Here and in the next figures, the various system parameters are expressed
    in units of $\kappa$. Moreover we use the following notation:
    ``UNI'' stands for uniform, ``CRY'' for crystalline, ``OSC'' for oscillatory,
    ``\dots/\dots'' denotes a bistability.
    If the pumping amplitude is further increased beyond $\Omega \gtrsim 1$,
    oscillations are no longer regular. Moreover, an instability sets in: two initial states
    with a very small difference would lead to a large difference in the time evolution.}
  \label{fig:Phase_U0}
\end{figure}

The stability of the fixed points should be analyzed as well.
If all the eigenvalues have negative real parts, the fixed point is stable.
If all the eigenvalues have negative real parts except for a pair of purely imaginary eigenvalues,
a Hopf bifurcation appears, thus we can expect to see a limit circle from the system.
The parameters $\Omega$ and $\delta$ can be controlled through the external driving
and are the easiest to be tuned in experiments (within the same array).
We thus start our discussion of the phase diagram as a function of these two parameters.
This is shown in Fig.~\ref{fig:Phase_U0}.
Here and in the next figures, we use the following notation:
``UNI'' stands for uniform, ``CRY'' for crystalline, ``OSC'' for oscillatory,
``\dots/\dots'' denotes a bistability.
A vanishing $\Delta n$ indicates the normal, uniform phase, while a non-zero $\Delta n$ signals
the crystalline phase in which the photon number is modulated as in a photon crystal.
Note that there is an oscillatory phase in the region
$0.8 \lesssim \delta \lesssim 1.3$ and $\Omega \gtrsim 0.9$, due to the
appearance of a Hopf bifurcation with increasing pumping amplitude.
In this phase the system state will never become completely stationary and,
in the long-time limit, the trace of $\langle a\rangle$
with $\langle a_A\rangle\ne\langle a_B\rangle$ is a limit circle.
Since, in our case, the Hopf bifurcation appears and disappears only for non-uniform fixed points,
$\Delta n$ will be different from zero in the oscillatory phase. 
A further investigation of the reduced density matrix
of the sublattice (either A or B) shows that the system is in a coherent state
(see Sec.~\ref{sec:Wigner} and~\ref{sec:squeezing} for more details).
The oscillatory phase also extends to finite values of $U$, although the coherent state
is progressively deformed on increasing the onsite repulsion.
The contemporary presence of checkerboard ordering and global dynamical phase coherence
suggested us to view this phase as a nonequilibrium supersolid phase~\cite{jin2013}.

Finally, let us also point out that two additional regions,
indicated with ``UNI/OSC'' and ``UNI/CRY'', are present
in the phase diagram of Fig.~\ref{fig:Phase_U0}. For the parameter values in these regions,
the steady state does depend on the initial values of the density matrix.
This indicates that the system is bistable.

\begin{figure*}[!t]
  \includegraphics[width = 0.33\linewidth]{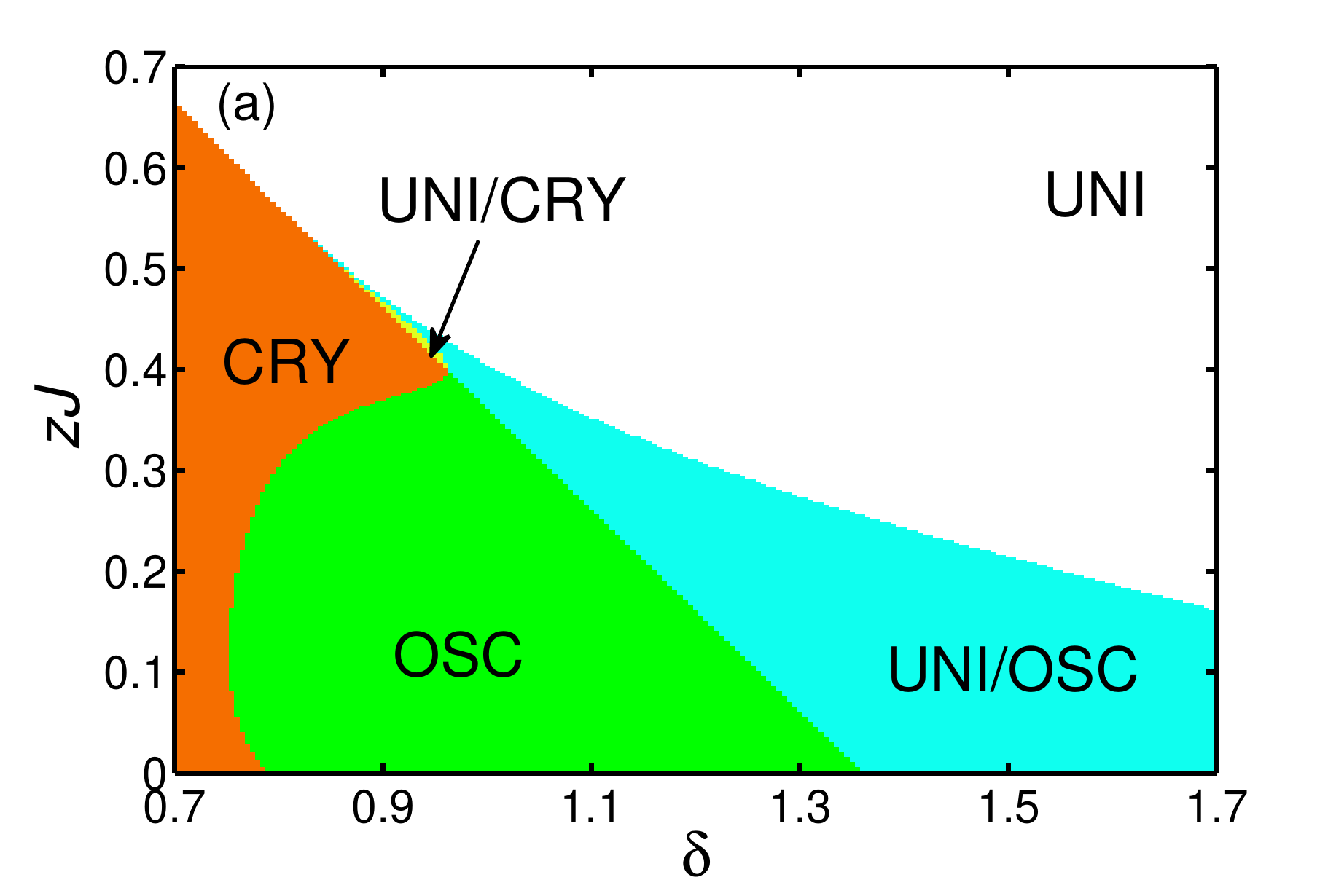}
  \includegraphics[width = 0.33\linewidth]{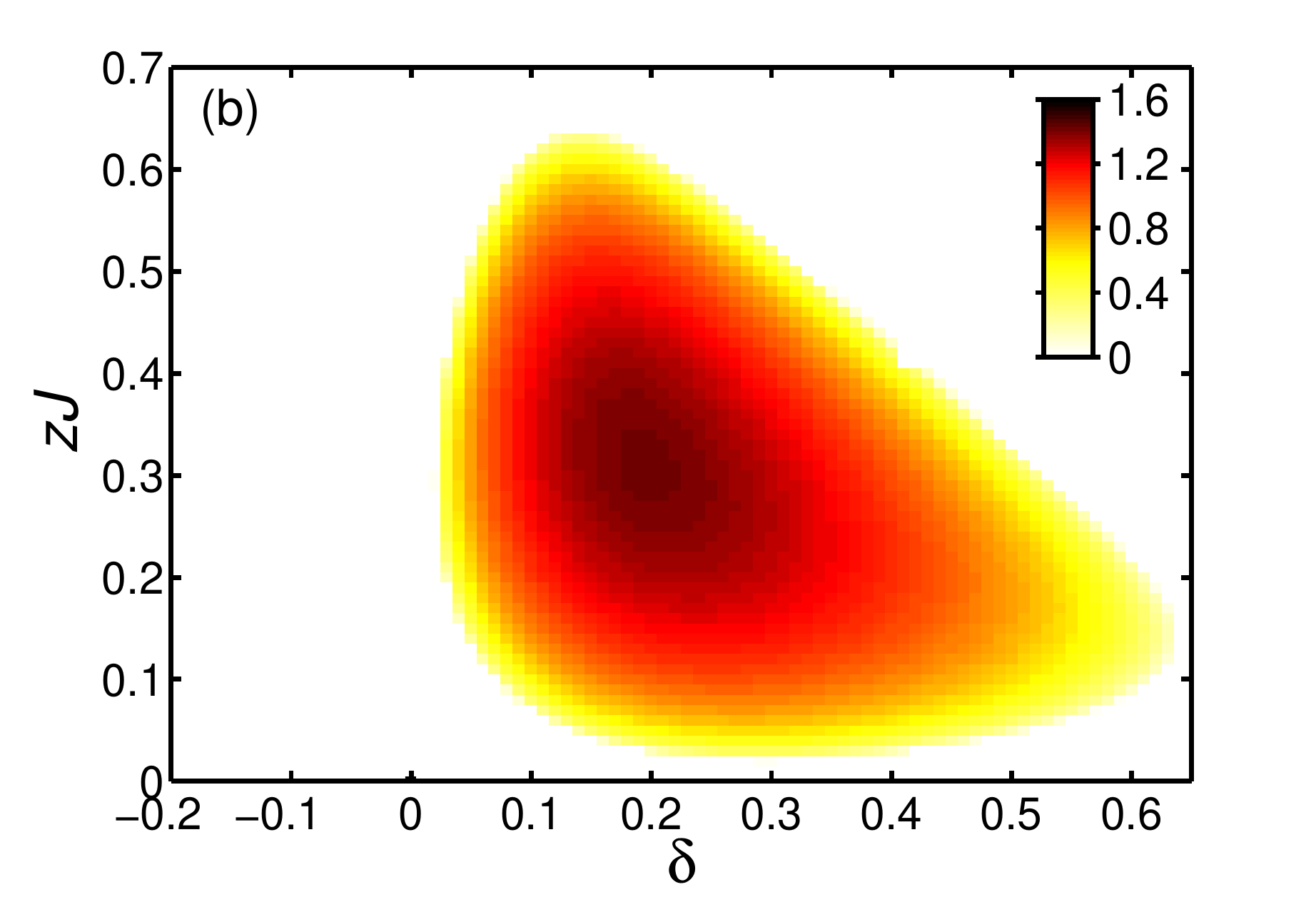}
  \includegraphics[width = 0.33\linewidth]{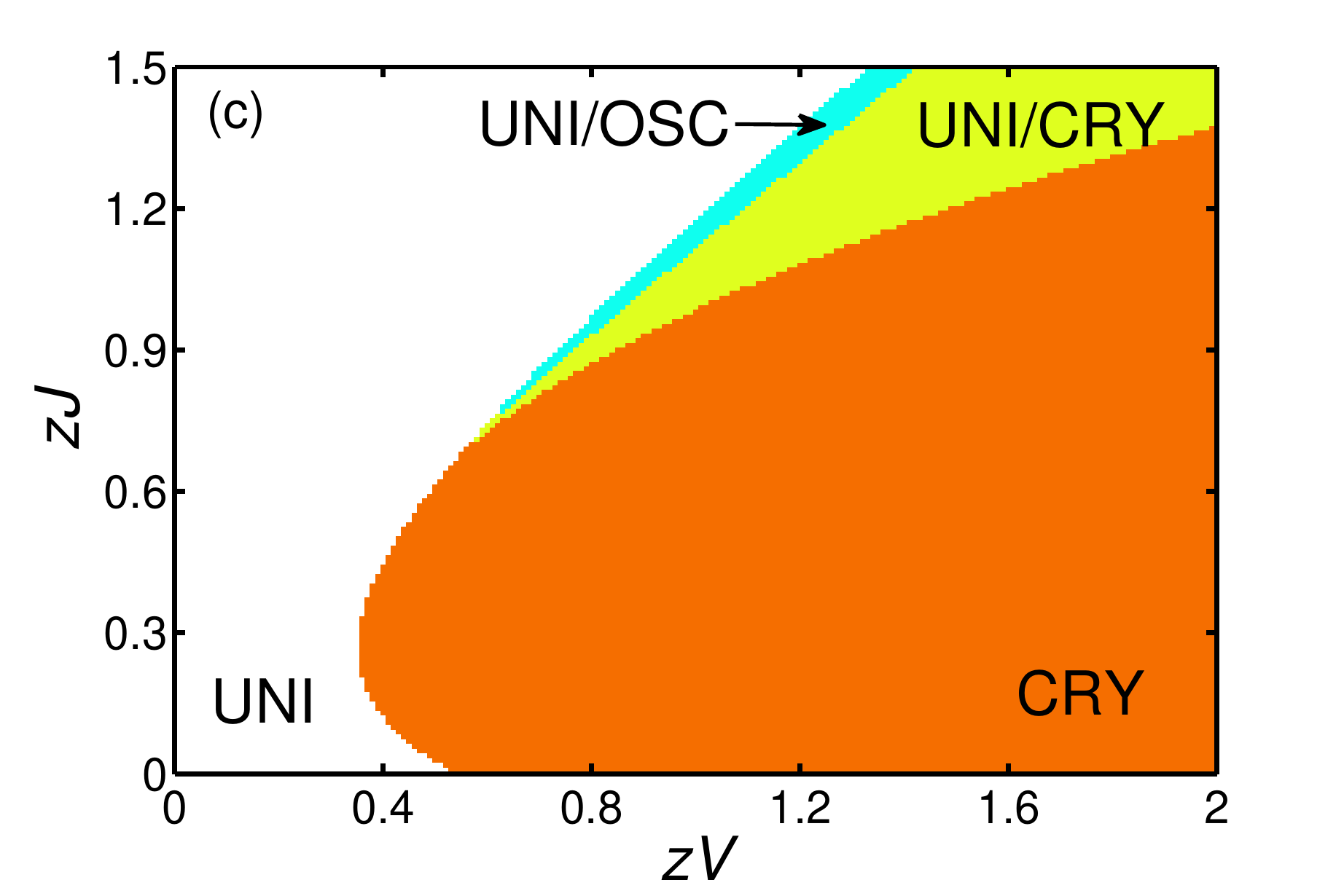}
  \caption{(Color online). Effects of a finite nearest-neighbor hopping $J$.
    Panel (a): Phase diagram in the $zJ$-$\delta$ plane, with $\Omega=1$.
    Panel (b): Crystalline order parameter $\Delta n = |\langle n_A\rangle-\langle n_B \rangle|$
    in the $zJ$-$\delta$ plane, with $\Omega = 0.6$. In this case the system exhibits
    either a uniform or a photon crystal phase.
    Panel (c): Phase diagram in $zJ$-$zV$ plane, with $\Omega=0.6$, and $\delta=0.2$.
    If not specified, the various parameters are set as in Fig.~\ref{fig:Phase_U0}.}
    \label{fig:Phase_J>0}
\end{figure*}

\subsubsection{Finite nearest-neighbor hopping}

In the case of $J\ne 0$,
it is possible to find the roots of Eq.~\eqref{dfeq} and check
the stability of the fixed points numerically. The phase diagram as a function
of the hopping strength $J$ is shown in Fig.~\ref{fig:Phase_J>0}, panel (a), where
we observe that the hopping delocalizes photons and favors the uniform phase.
Together with the quench of the crystalline phase, finite values of $J$
hopping may facilitate a crystalline order,
thus leading to a reentrance in the phase diagram [panel (b)].
A qualitatively similar feature can be also seen in the $J-V$ plane, as shown
in panel (c). At this stage there is no simple explanation for the reentrance.
The fact that the hopping may stabilise the crystalline phase indicates that
quantum fluctuations are important.
Moreover we would like to stress that the reentrance might also appear 
as a peculiarity of the mean-field approximation.

\begin{figure}[b]
  \includegraphics[width = 1\linewidth]{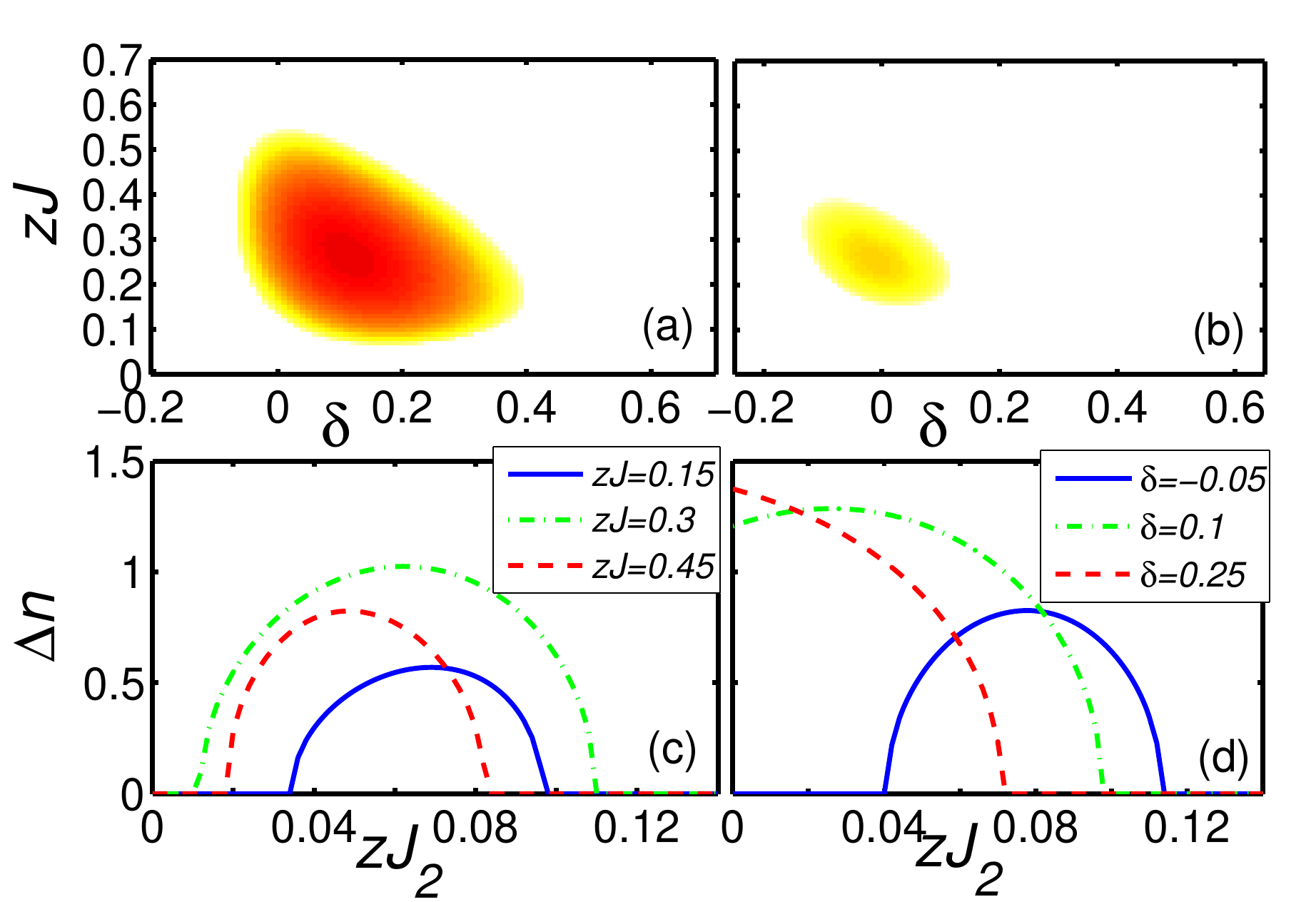}
  \caption{(Color online). Crystalline order parameter in presence of a correlated hopping.
    For simplicity we choose $J_2=J_n$.
    The other parameters are $U = 0$, $zV = 0.5$, and $\Omega = 0.6$.
    Upper panels: we fix (a) $zJ_2=0.05$ and (b) $zJ_2=0.1$,
    and show $\Delta n$ in the $zJ$-$\delta$ plane
    [hereafter we use the same color scale as in Fig.~\ref{fig:Phase_J>0}(b)].
    For the sake of a comparison with the case in absence of correlated hopping,
    we used the same scale as in Fig.~\ref{fig:Phase_J>0}(b).
    Lower panels: $\Delta n$ as a function of $J_2$ (c) fixing $\delta = 0.2$
    and for different values of $J$;
    (d) fixing $zJ=0.4$ and for different values of $\delta$.}
  \label{fig:Phase_J2>0}
\end{figure}

\subsubsection{Finite correlated hopping}

The effect of a finite correlated and pair hopping
($J_2 \neq 0$, $J_n \neq 0$) is illustrated in Fig.~\ref{fig:Phase_J2>0}.
For simplicity we chose $J_2=J_n$ (changing this ratio introduces only quantitative differences).
As for the nearest-neighbor hopping, pair hopping generally increases
the extension of the uniform phase.
In particular, compare the upper panels of Fig.~\ref{fig:Phase_J2>0}
with the corresponding panel (b) in Fig~\ref{fig:Phase_J>0}, where we observe
that, for larger values of $J_2$ and $J_n$, the colored region shrinks
and shifts towards smaller $\delta$ values.
Apart from some quantitative modifications, however the shape of the phase diagram is not modified,
thus confirming what has been anticipated in Ref.~\cite{jin2013}.

\begin{figure}[b]
  \includegraphics[width = 1\linewidth]{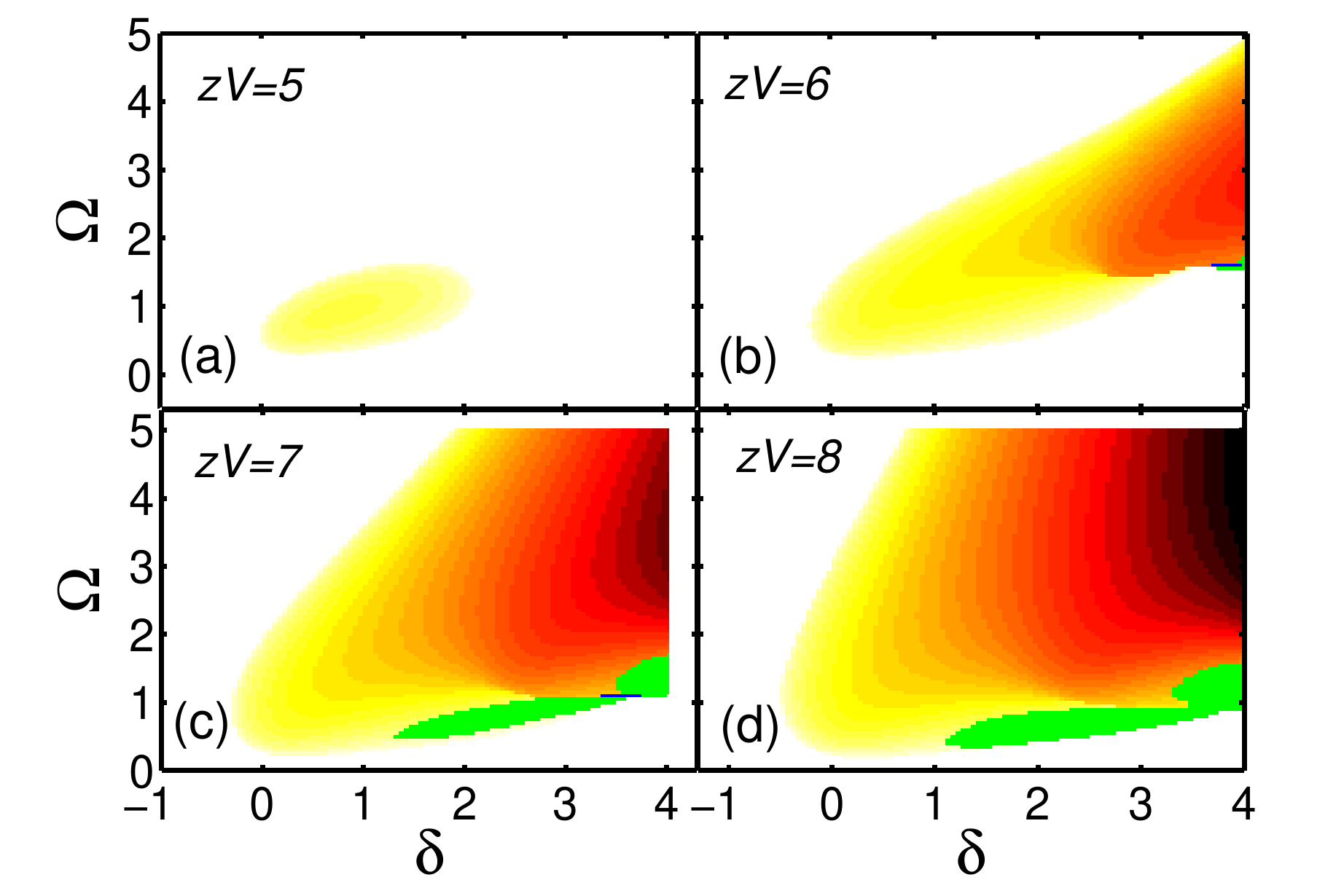}
  \caption{(Color online). Effects of a finite onsite interaction:
    phase diagram in the $\Omega$-$\delta$ plane
    for $U=2$, different values of $V$, and zero hopping $J = J_2 = J_n = 0$.
    The tiny regions in green and blue denote oscillatory and ``CRY/OSC''
    bistability phases, respectively.
    An increasing cross-Kerr nonlinearity extends the area of crystalline phase
    and leads to the appearance of oscillatory phases.}
  \label{fig:FiniteU}
\end{figure}

In the bottom panels of Fig.~\ref{fig:Phase_J2>0}, the effect of the correlated/pair hopping
is further analyzed.
It is evident that, for very small values, a non-zero value of $J_2$ contributes to stabilize
the crystal phase. In this regime, the crystalline phase is already quenched by quantum fluctuations
and correlated hopping may be an efficient means to homogenize those configurations
with higher occupation that do not contribute to the order.
Eventually continuing to increase $J_2$, there is a transition back to a homogeneous phase.
The same effect is also seen in panel (d), where the different curves are parametrized by the detuning.

\subsection{Finite onsite interaction ($0 < U < \infty$)}

In the case in which the onsite repulsion is taken into account,
the mean-field equations have to be solved numerically.
The results for different values of $V$ and non-zero onsite repulsion are summarized
in Fig.~\ref{fig:FiniteU}, in the limit of zero photon hopping.
The cross-Kerr nonlinearity $V$ tends to extend the crystalline phase.
For some intermediate value of $\Omega$ and $\delta$, an oscillatory phase emerges.
However due to the onsite nonlinearity the reduced density matrix of the sublattice $A$ or $B$
is not coherent any more. We will discuss this issue further in the remaining part of the paper.

\begin{figure}
 \begin{minipage}[b]{8.5cm}
 \includegraphics[width = 0.98\linewidth]{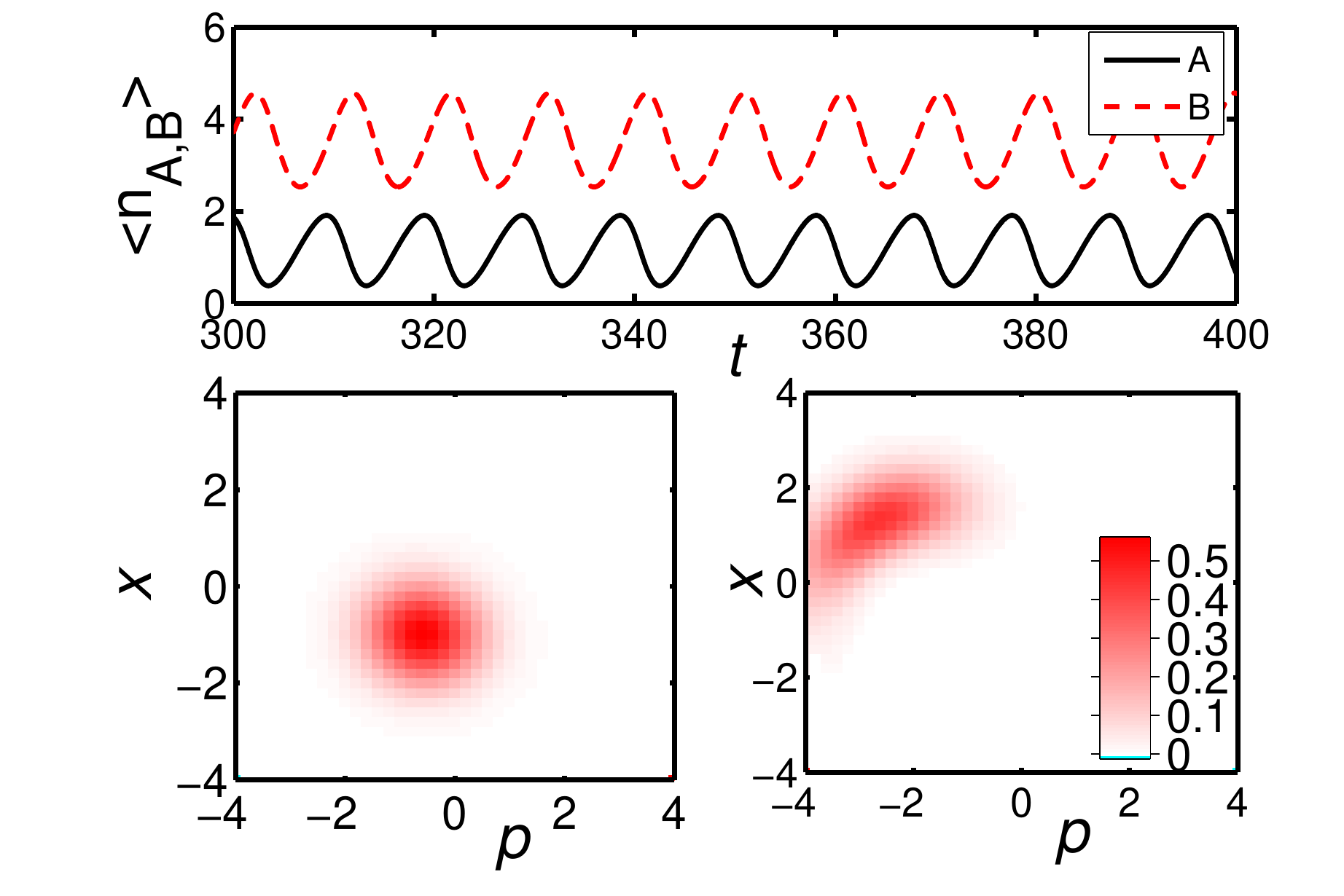}
  \vspace{0.6cm}
 \end{minipage}
 \begin{minipage}[b]{8.5cm}
  \includegraphics[width = \linewidth]{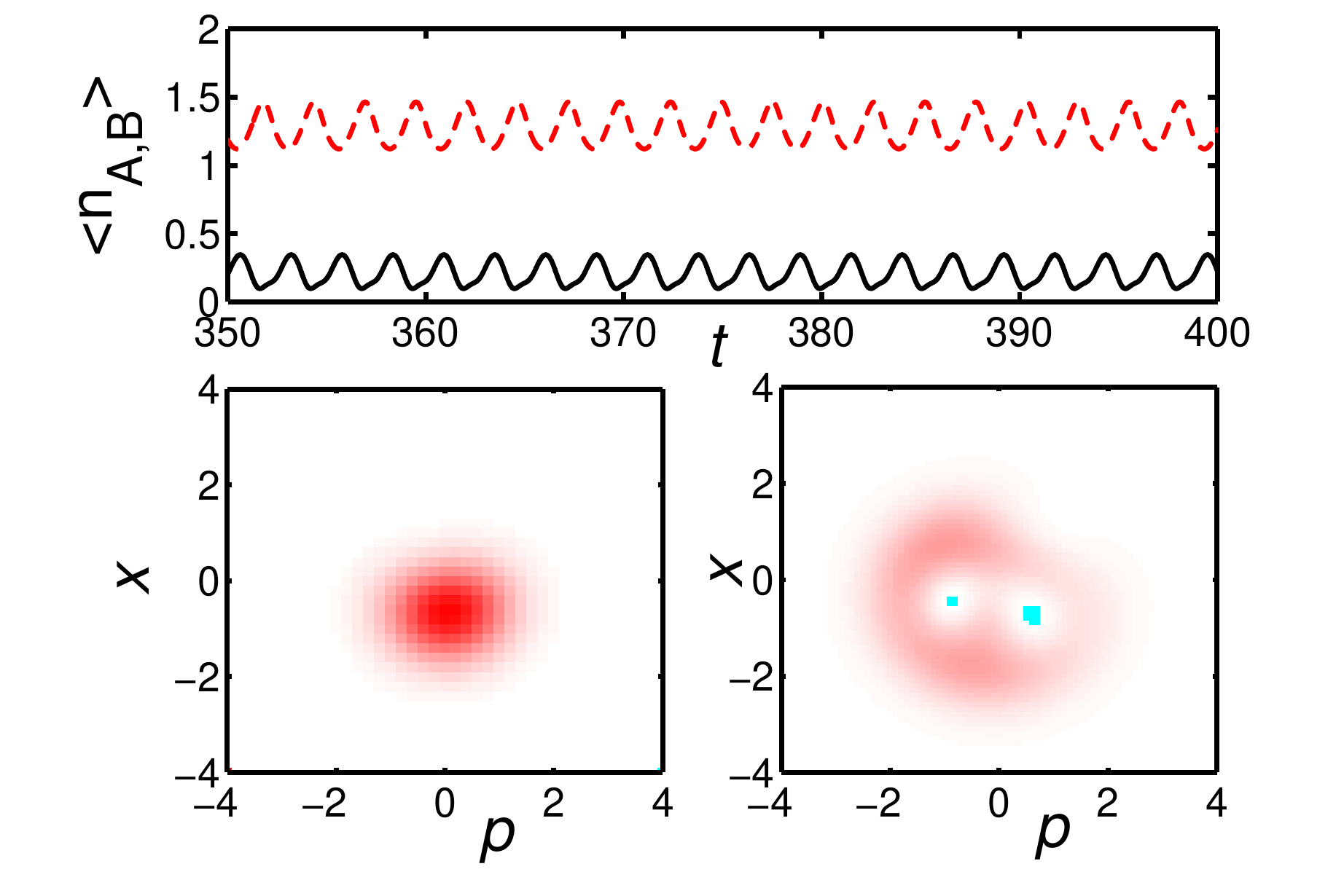}
  \vspace{-0.3cm}
  \caption{(Color online). Time evolution of $\langle n_{A,B} \rangle$ and corresponding
    Wigner functions for the two sublattices $A$ (solid black line and square panels on the left) and
    $B$ (dashed red line and square panels on the right) in the long-time limit at $t=400$,
    for various choices of the system parameters. Here we set $J_2=J_n=0$.
    The upper panels are in the small-$U$ regime: $zJ=0.2$, $U=0.05$, $zV=0.5$, $\Omega=1$, $\delta=1.2$.
    The bottom panels are in the large-$U$ regime: $J=0$, $U=2$, $zV=8$, $\Omega=2$, $\delta=6$.
    Cyan areas denote regions in which $W(x,p)$ becomes negative,
    even if it stays very close to zero in magnitude ($-0.02 \lesssim W \lesssim 0.6$).
    Here and in Fig.~\ref{fig:Wigner2}, times are chosen in units of $\kappa^{-1}$.}
  \label{fig:Wigner}
 \end{minipage}
\end{figure}

\subsection{Oscillatory and bistable phases -- Wigner function}
\label{sec:Wigner}

As we already mentioned, there are regions of the steady-state phase diagram
where the system shows a limit circle or a bistable behavior.
Let us have a closer look at these cases.
Here we will mostly concentrate on the case in which the onsite nonlinearity is present,
since the limit $U=0$ was already discussed in Ref.~\cite{jin2013}.

The rectangular panels in Fig.~\ref{fig:Wigner} display the time evolution
of the photon number $\langle n_{A,B} \rangle$ in each sublattice,
in the small-$U$ and in the large-$U$ regime.
We address directly the long-time limit, where we ensured that the system
has already reached the asymptotic state (this does not depend on the choice of the
initial conditions).
In the square panels we show the corresponding Wigner functions of the two sublattices
(A on the left, B on the right).
They are defined as
\begin{equation}
  W(x,p) = \int_{-\infty}^{\infty} \langle x-y |\rho_{A,B}| x + y \rangle e^{2ipy} {\rm d}y
\end{equation}
with $x=(a+a^\dagger)/\sqrt{2}$, $p=i(a^\dagger-a)/\sqrt{2}$, $|x\rangle$ being
an eigenstate of the position operator $x$ and $\rho_{A,B}$ being the reduced density matrix
of sublattice A or B.

\begin{figure}[!t]
  \includegraphics[width = 1.\linewidth]{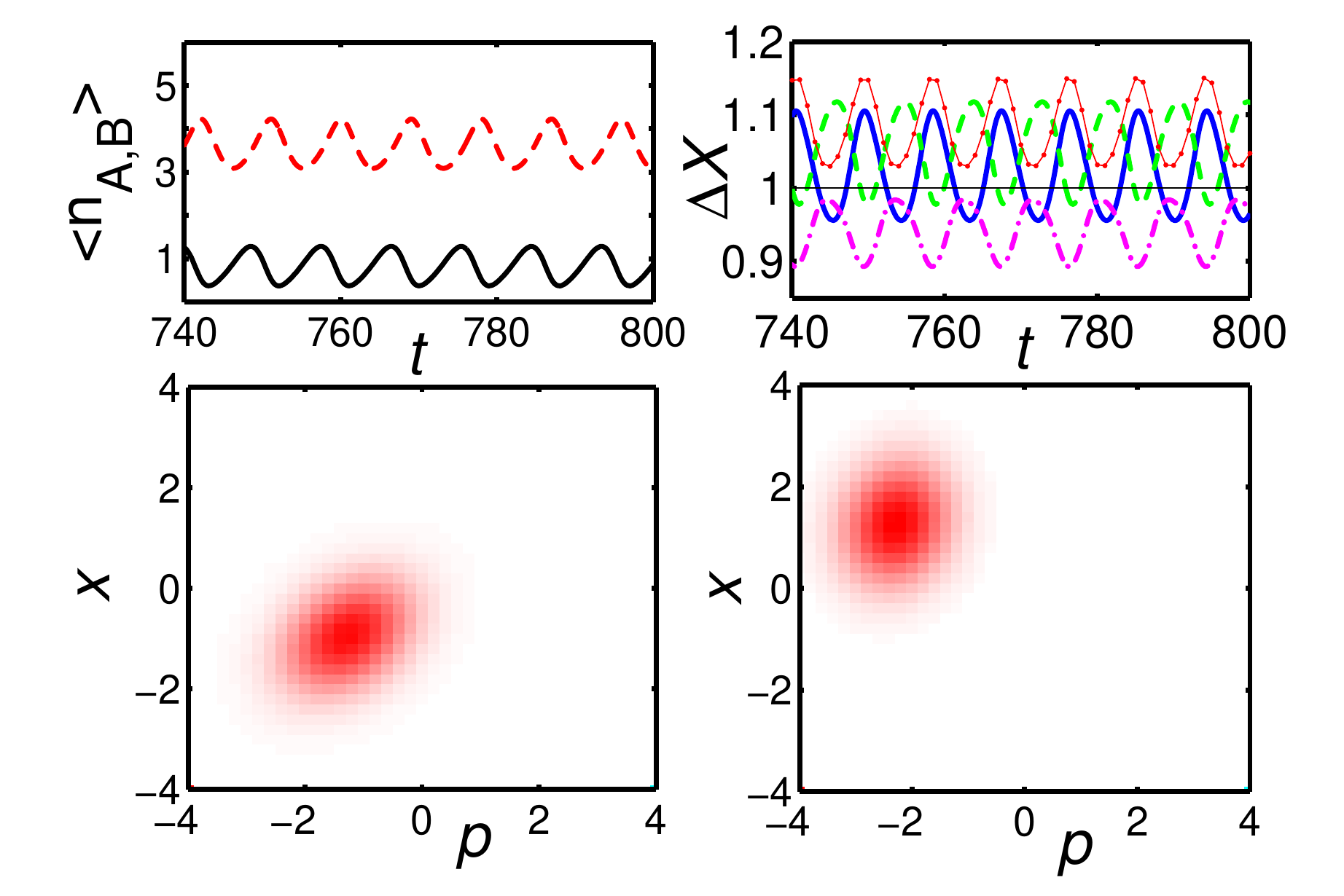}
  \caption{(Color online). Pair-hopping effects in the Wigner-function
    analysis of the two sublattices (similar plots as in Fig.~\ref{fig:Wigner}).
    Here we set $U=J_n=0$, $zJ=0.3$, $zJ_2=0.1$, $zV=0.5$, $\Omega=1$, $\delta=0.6$.
    The top right panel displays squeezing properties in the two sublattices
    [continuous/dashed lines --- see caption of Fig.~\ref{fig:Squeezing} (c)].}
  \label{fig:Wigner2}
\end{figure}

For small onsite nonlinearity $U$ (three upper panels),
the system evolves periodically in time and the asymptotic state is a limit circle.
Compared with the limiting case of $U=0$, the coherence of each sublattice
is drastically modified due to the presence of the onsite repulsion.
This can be seen from the Wigner function of each sublattice: it is clear that
the distributions in phase space deviate from the Gaussian shape,
especially for the sublattice with higher photon number.
Furthermore, on increasing the interaction strength, coherence is progressively weaker
and correspondingly synchronization is suppressed.
For the oscillatory regions at large $U$ (three bottom panels), the Wigner function
of the sublattice with higher photon number may resemble a two-hole ringlike shape,
which means that the phase of the motion is undetermined.
On the contrary, for small $U$ the system is still synchronized, albeit not perfectly
because there may be a (small) error in the determination of the phase.

Also in the presence of correlated hopping (Fig.~\ref{fig:Wigner2}), oscillatory phases do appear.
From Eq.~\eqref{Eq_meanfield}, we see that the pair hopping is likely to introduce
a squeezing effect to the mode of each sublattice in which the uncertainty of one
variable is reduced by sacrificing the certainty of the conjugate one.
In order to see the squeezing properties of the system, we may write the annihilation
operator $a$ as a linear combination of two Hermitian operators, $a=(X_1+iX_2)/2$ with
the operators $X_1$ and $X_2$ obeying $[X_1,X_2] = 2i$.
The corresponding uncertainty relation is $\Delta X_1 \, \Delta X_2 \ge 1$,
where $\Delta X_{j}=\sqrt{\langle X_{j}^2\rangle-\langle X_{j}\rangle^2}$ ($j=1,2$).
For a coherent state we always have $\Delta X_1=\Delta X_2=1$, while for a squeezed state
$\Delta X_1 < 1 < \Delta X_2$, so that the uncertainty of one quadrature
is reduced at the expense of increasing the uncertainty of the other one.
It is therefore tempting to associate the different regions of the phase diagram
to different squeezing behavior.

\begin{figure}[t]
  \includegraphics[width = 1\linewidth]{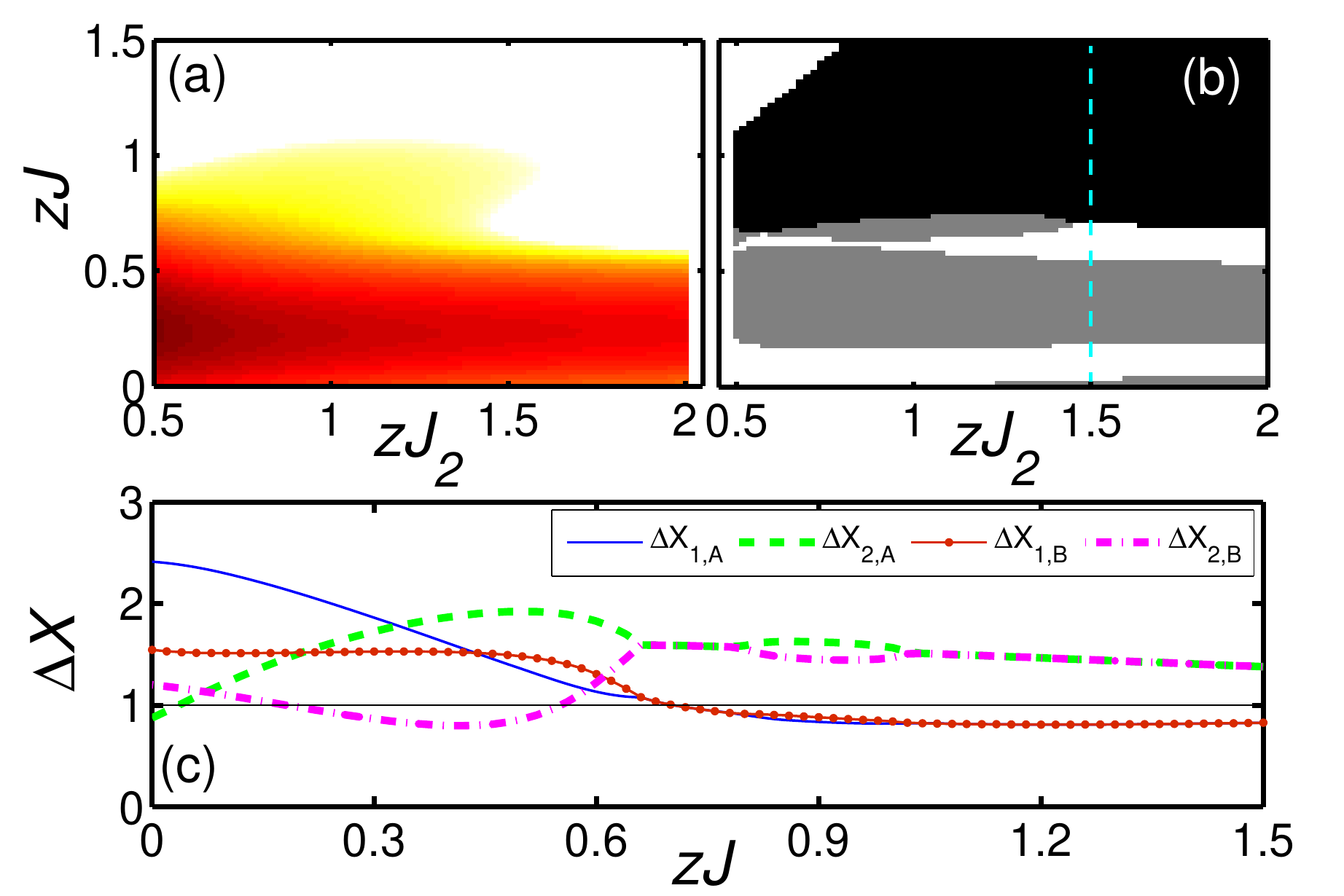}
  \caption{(Color online). Squeezing of the steady state in the $zJ-zJ_2$ plane,
    for $U=0$, $zV=0.5$, $\Omega=1$, and $\delta=0.8$.
    The crystalline order parameter $\Delta n$ [panel (a)] is compared
    with the squeezing properties [panel (b)] in the same parameter space.
    The system can exhibit squeezing of both two (black),
    only one (gray), or none (white region) of the sublattices.
    Panel (c): variance of operators $X_1$, $X_2$ for the two sublattices,
    plotted respectively with continuous and dashed lines, as a function of $zJ$,
    for $zJ_2=1.5$ [{\it i.e.}, along the vertical line on panel (b)].
    Squeezing is signaled by values of $\Delta X < 1$ (horizontal straight line, as a guide to the eye).}
  \label{fig:Squeezing}
\end{figure}

Looking carefully at the upper right panel of Fig.~\ref{fig:Wigner2}, we note that,
for our choice of the parameters, sublattice B (dashed lines) is always squeezed.
On the other hand, the squeezing property of sublattice A (continuous lines) is time-dependent.
The bottom panels display the Wigner functions at the moment in which
both sublattices are squeezed.

\subsection{Correlated hopping and squeezing}
\label{sec:squeezing}

A further example of the squeezing is shown in Fig.~\ref{fig:Squeezing},
where we analyzed its behavior in the $J$-$J_2$ parameter space,
in absence of the onsite interaction, $U=0$.
The squeezing properties can be divided into three regions.
In the black region both of the two sublattices are squeezed, in the gray region
only one sublattice is squeezed, and in the white region no sublattice is squeezed.
A more detailed analysis based on the variance is shown in panel (c).

\section{Cluster mean-field approximation}
\label{clusterm}

The single-site mean-field approximation ignores all quantum correlations
between the subsystems~\cite{Degenfeld13}.
In order to get a flavor of the role of correlations, we employ a (more demanding)
cluster mean-field approach. In this case, short range correlations within the cluster are treated
exactly, while the mean field is defined at the boundary of the cluster itself.
Our hope is to have more accurate information on the phase diagram,
because short range correlations are preserved.

\begin{figure}[b]
  \includegraphics[width = 0.7\linewidth]{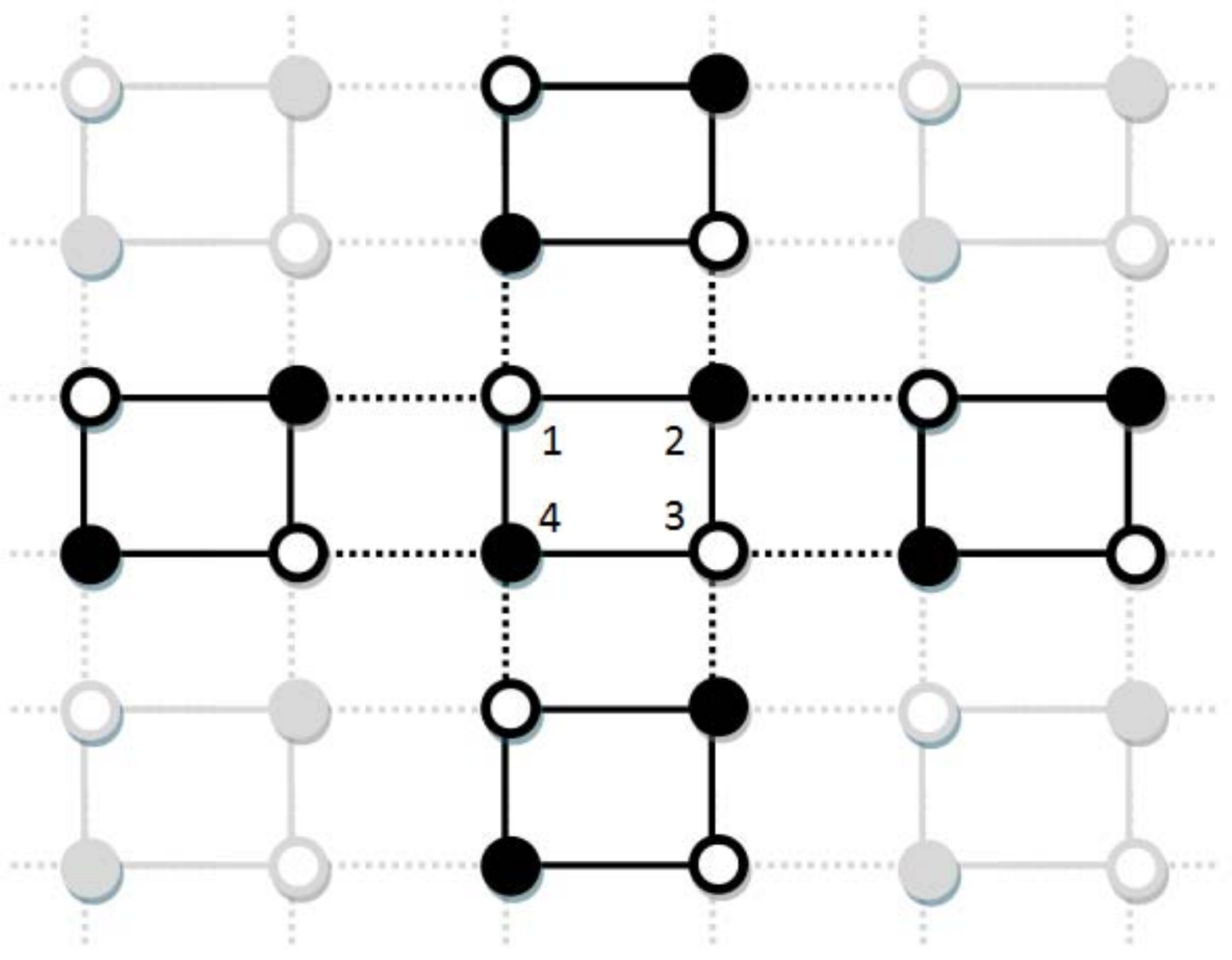}
  \caption{Schematic representation of a two-dimensional lattice in terms of square clusters.
    Each cluster (solid square) is composed of two $A$ sublattices and two $B$ sublattices.
    The interactions within the cluster (solid lines) are computed exactly.
    Interactions between two neighboring clusters (dotted lines) are treated as mean fields.
    The white and black circles denote sublattice $A$ and $B$, respectively.}
  \label{cluster}
\end{figure}

We divide the lattice into clusters composed of four sites (clockwise labelling
the sites by 1, 2, 3, and 4). Without loss of generality, we identify the sites with
odd (even) numbers with sublattices $A$ ($B$).
For the sake of clarity, let us focus on a two-dimensional
square lattice in this Section, see Fig.~\ref{cluster}.

The Hamiltonian of each cluster can be exactly written according to the following
\begin{eqnarray}
  {\cal H}_{\rm C} & = & \sum_{i=1}^{4} \left[ -\delta n_i + \Omega(a_i + a_i^\dagger) + U n_i(n_i-1) \right] \cr\cr
  & & + V\sum_{\langle i,j \rangle} n_i n_j - J\sum_{\langle i,j \rangle} (a_i^\dagger a_j + \text{H.c.}) \cr\cr
  & & + \sum_{\langle i,j \rangle} \left[ \frac{J_2}{2} a_i^\dagger a_i^\dagger a_j a_j
      - J_n a_i^\dagger(n_i+n_j)a_j+\text{H.c.} \right] \, .
\end{eqnarray}
The interactions between neighboring clusters are treated at the mean-field level,
and thus the coupling of the cluster with the rest of the system is described by the (mean-field) term,

\begin{figure}[t]
  \includegraphics[width = 1\linewidth]{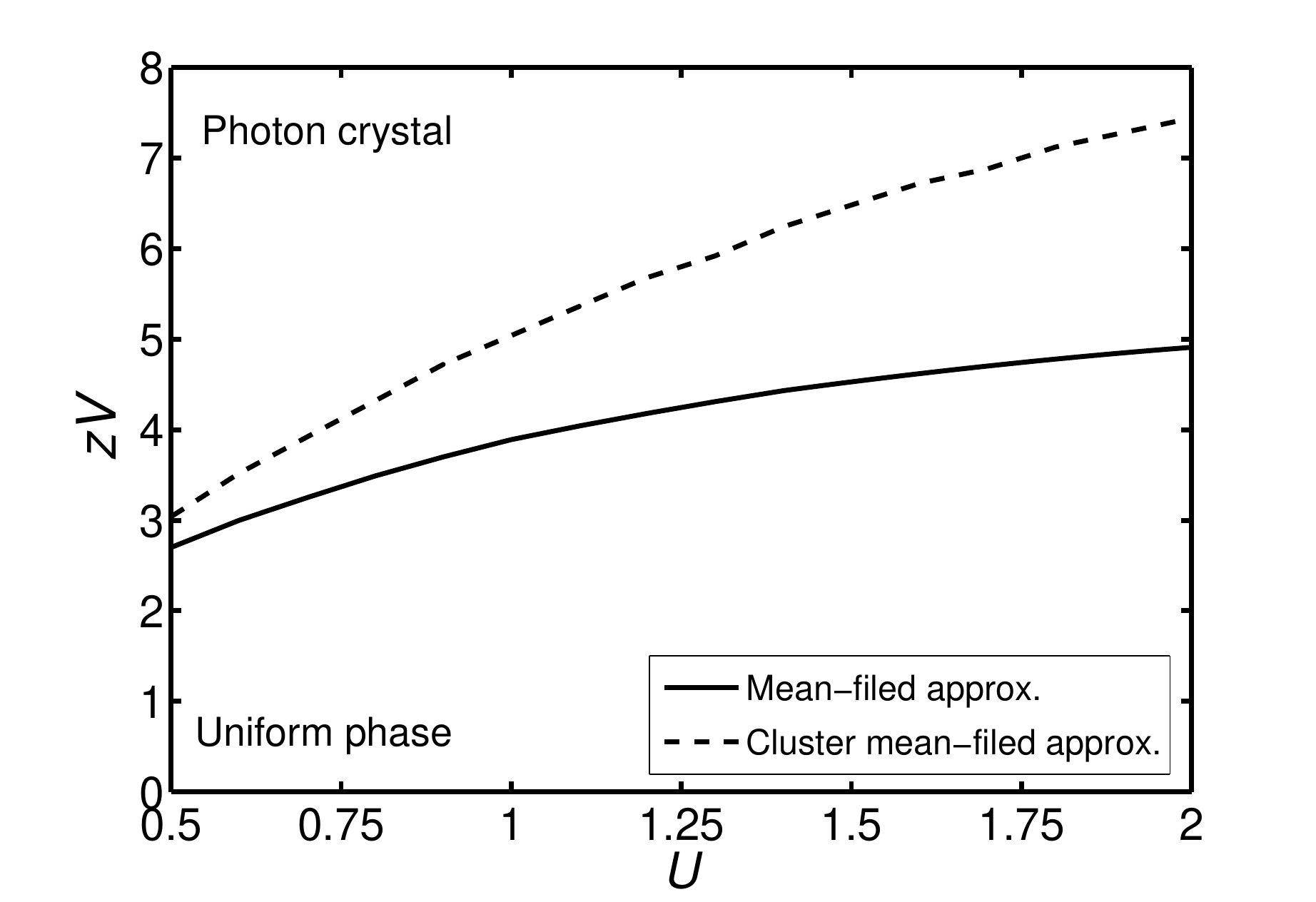}
  \caption{Phase diagram in the $U$-$V$ plane obtained with mean-field and cluster
    mean-field approximations, in absence of hopping ($J = J_2 = J_n = 0$),
    and for $\delta = 0$, $\Omega = 0.75$.
    The curves separate regions where the system is
    in a crystalline (above) and in a uniform (below) phase.
    Here we set $z=4$, thus resembling a two-dimensional lattice.}
  \label{phase10}
\end{figure}

\begin{eqnarray}
  {\cal H}_{\rm C-MF} & = & 2 V \sum_{i\in A,B} \langle n_{\bar{i}} \rangle n_i
  - 2 J \sum_{i\in A,B} \left[ \langle a_{\bar{i}} \rangle a_i^\dagger + \text{H.c.} \right] \cr\cr
  & & + J_2\sum_{i\in A,B} \left[ \langle a_{\bar{i}}^2 \rangle a_i^{\dagger 2}+ \text{H.c.} \right] \cr\cr
  &&-2J_n\sum_{i\in A,B} \left[ \langle n_{\bar{i}}a_{\bar{i}} \rangle a_i^\dagger +\langle a_{\bar{i}}\rangle a_i^\dagger n_i+ \text{H.c.} \right] \, ,
\end{eqnarray}
where $\bar{i}$ labels a site in the sublattice different from that to which the $i$-th site belongs. 
Note that also in the cluster-mean field the symmetry is explicitly broken once the cluster 
is coupled to the rest of the lattice through ${\cal H}_{\rm C-MF}$.

For the case of $J = J_2 = J_n = 0$, the phase diagram in the $U-V$ plane, obtained via mean-field
and cluster mean-field approximations, is shown in Fig.~\ref{phase10}.
Here again we use the order parameter $\Delta n = |n_A-n_B|$ to distinguish
the crystalline and uniform phases. If the cross-Kerr term exceeds a critical
threshold $z V_c$, the steady state is characterized by a staggered order in which $\Delta n \ne 0$.
It can be seen that in the cluster mean-field case, the crystalline phase is reduced.
For small values of $U$ the discrepancy between the two approaches is tiny, 
but it increases for larger $U$.
In the hard-core limit ($U \rightarrow \infty$) the critical point obtained
in the cluster mean-field approximation, $z V_c^{\rm (C-MF)} \approx 11.76$,
is about twice larger than that in the mean-field approximation, $zV_c^{\rm (MF)} \approx 5.73$.

\begin{figure}[t]
  \includegraphics[width = 1\linewidth]{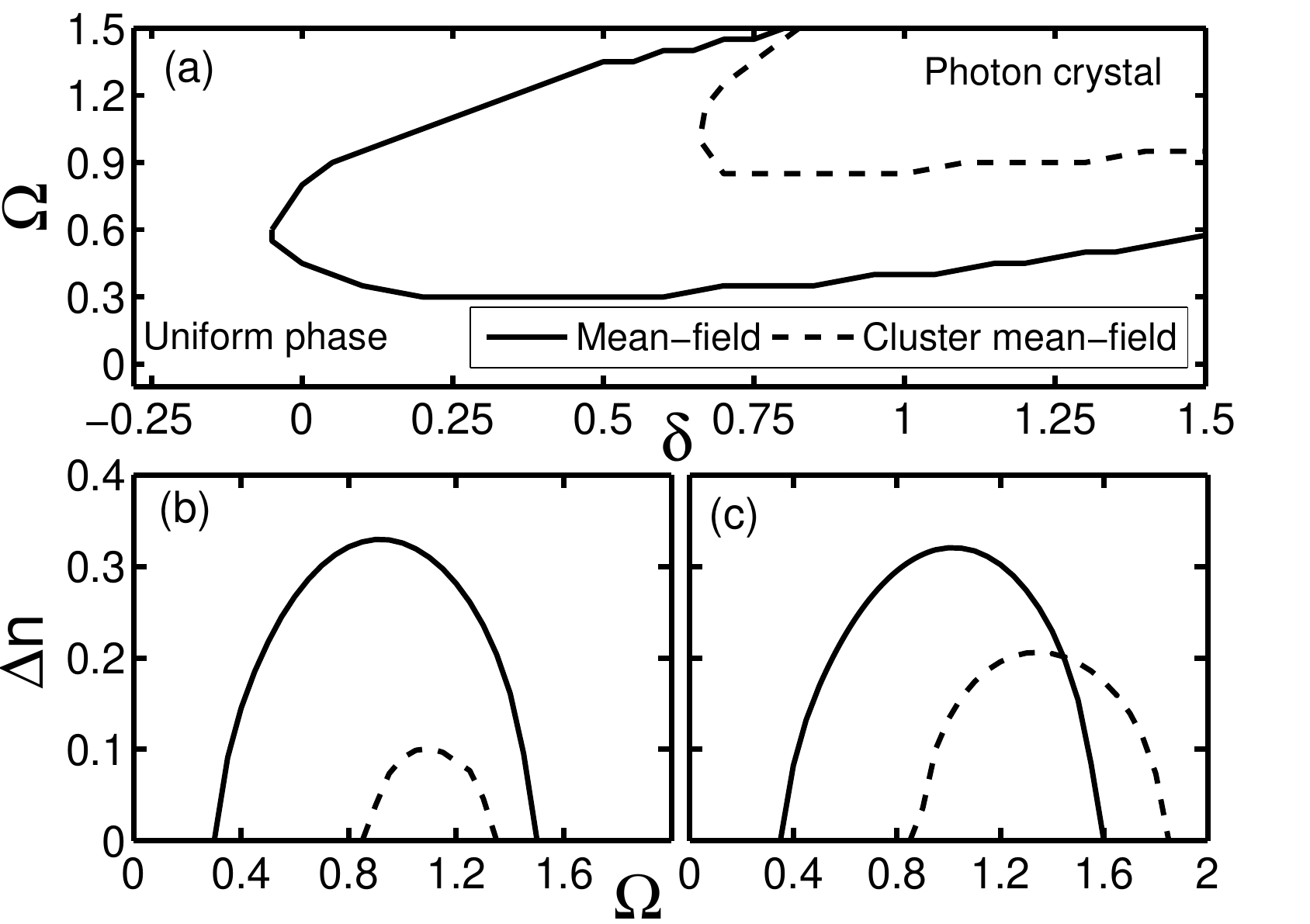}
  \caption{Panel (a): Phase diagram in the $\Omega$-$\delta$ plane obtained with mean-field 
    and cluster mean-field approximations, in absence of hopping ($J = J_2 = J_n = 0$),
    and for $U = 2$, $zV = 5$.
    The curves separate regions where the system is in a crystalline (inside) 
    and in a uniform (outside) phase. 
    Panels (b-c): Order parameter $\Delta n$ computed with mean-field 
    and cluster mean-field approximations, as a function of $\Omega$,
    for $\delta = 0.75$ and $\delta = 1$, respectively.}
  \label{phase11}
\end{figure}

We further investigate the role of short-range quantum fluctuations, 
taken into account by the cluster mean field, by considering the onset 
of the crystalline phase as a function of the driving $\Omega$ and of 
the detuning $\delta$.
The results of this analysis are presented in Fig.\ref{phase11}
[for a direct comparison, see the analogous calculation with 
the single-site mean field shown in Fig.~\ref{fig:FiniteU}, panel (a)]. 
As for the study in the $U-V$ plane, taking into account the short-range 
fluctuations leads to a shrinking of the extension of the photon crystal,
at least in the region $0 \leq \Omega \leq 1.5$, $0 \leq \delta \leq 1.5$. 
Our simulations suggest the persistence of the crystalline phase even
for higher values of the detuning, where a much larger photon number per cavity 
has to be considered, thus greatly enhancing the computational effort.
Despite all these quantitative modifications, the results of the single-site 
analysis seem to be quite robust.

On a broader perspective it would be very interesting to see, through simulations 
on small clusters, to which extent a signature of the results presented here 
can be seen in few cavity systems. 
In Ref.~\cite{jin2013} simulations on cavity chains confirmed the onset 
of (short-range) ordering. It would be very important to confirm it 
also for "two-dimensional" clusters.

\section{Circuit-QED cavity array with nonlinear couplings}
\label{cQED}

So far we analyzed the generic problem of a driven cavity array with nonlinear coupling.
In this last Section we analyze in more detail the emergence of the crystalline phase 
in an implementation with circuit-QED arrays. In this case not all the coupling constants 
can be chosen freely.

Circuit-QED is particularly well suited for implementing nonlinear couplings
between cavities or resonators, because of its great design flexibility,
the dissipation-less nonlinearity provided by Josephson junctions, and the exceptionally
high coupling between neighboring elements that can be reached.
Here the latter can be mediated via a Josephson junction.
As already discussed in Ref.~\cite{jin2013}, our goal is to realize a cavity array
with a strong cross-Kerr nonlinearity. This can be achieved, for example,
using the circuit depicted in Fig.~\ref{crossKerrSetup}.
This scheme has been already described in the Supplementary Information of Ref.~\cite{jin2013},
and here we recap the main ingredients of this implementation in order to make the paper self-contained.

The building block of the cavity array is shown in Fig.~\ref{crossKerrSetup},
where adjacent cavities, labeled as sites $i$ and $i+1$, are coupled via a Josephson junction.
We focus on lumped element resonators (see Fig.~\ref{crossKerrSetup}a for a sketch) 
to keep the derivation simple and transparent.
Coplanar waveguide resonators work equally well (Fig.~\ref{crossKerrSetup}b).
In the following we concentrate on the building block of the nonlinear coupled array and discuss the
nonlinearity in the coupling of two cavities.
We ignore any on-site nonlinear circuits, since this can be added in the standard
way~\cite{wallraff2004}, by coupling each resonator (LC-circuit) locally to an additional qubit.

\begin{figure}
  \includegraphics[width=\columnwidth]{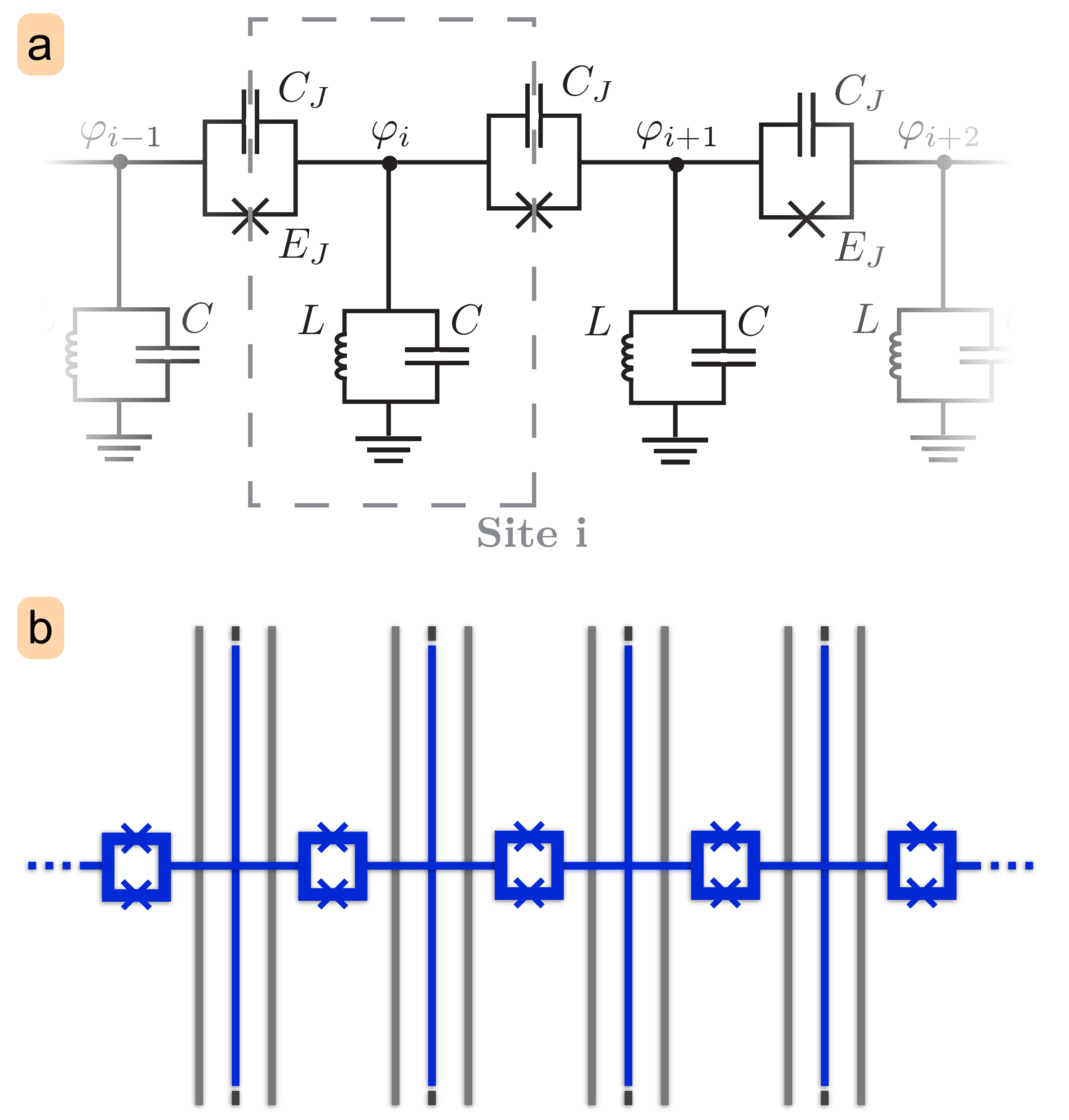}
  \caption{(Color online). Electrical circuit sketch of the setup we envision to realize a system
    with cross-Kerr nonlinearities. {\bf a}) Lumped element version with LC-circuits, 
    representing the cavities, in one dimension.
    Adjacent LC-circuits are coupled via tunable, capacitively shunted Josephson junctions 
    (each formed by a dc-SQUID with two junctions in parallel).
    {\bf b}) Realization with coplanar waveguide resonators. 
    The central conductors of adjacent resonator are connected with a conductor 
    that is intersected by the dc-SQUID forming the nonlinear coupling element (blue).
    Here we draw $\lambda$-resonators with a current anti-node in the center. 
    The capacitive in- and output ports remain accessible for drives and measurements.}
  \label{crossKerrSetup}
\end{figure}

In terms of the node fluxes $\varphi_i$, the Lagrangian of the two-cavity system reads,
\begin{equation}
  \mathcal{L} =
  \sum_{i=1,2} \left[\frac{C}{2} \dot{\varphi}_i^2 - \frac{1}{2 L} \varphi_i^2\right]
  + \frac{C_J}{2} \dot{\varphi}_{12}^2 + E_J \cos\left(\frac{\varphi_{12}}{\varphi_{0}}\right) \,.
\end{equation}
In this expression $L$ and $C$ are respectively the inductance and capacitance
of the lumped element resonators, $C_J$ and $E_J$ the capacitance and Josephson energy
of the Josephson junctions introduced for the coupling between the cavities,
$\varphi_{0} = \hbar/(2 e)$ the reduced quantum of flux,
and $\varphi_{12} = \varphi_{1} - \varphi_{2}$.
The corresponding Hamiltonian can be derived~\cite{Devoret_1995} by introducing
the charges on the islands $q_i$, canonically conjugated to the fluxes $\varphi_i$.
The quantized form is then obtained by means of bosonic lowering
and raising operators $a_i$ and $a_i^{\dag}$ that relate to $\varphi_i$ and $q_{i}$ via
$\varphi_i = \big( \tilde{L}/4\tilde{C} \big)^{1/4} \, (a_i+a_i^{\dag})$ and
$q_i = i \big( \tilde{C}/4\tilde{L} \big)^{1/4} \, (a_i^{\dag}-a_i)$ with $\tilde{C}=C+2C_J$,
$1/\tilde{L}=1/(2L)+1/L_J$ and $L_{J} = \varphi_0^2 / E_J$.

Expanding the nonlinearities $\cos(\varphi_{12}/\varphi_{0})$ up to fourth-order 
in $\varphi_{12}/\varphi_{0}$ and performing a rotating wave approximation,
we arrive at the effective Hamiltonian,
\begin{equation}
  {\cal H} = {\cal H}_{lc} + {\cal H}_{os} + {\cal H}_{ck} + {\cal H}_{ch} \,,
  \label{hamiltonian-cQED}
\end{equation}
where
\begin{eqnarray}
  {\cal H}_{lc} \!& = \!& \omega X_{J}(a_1^{\dag}a_{2}+a_1a_{2}^{\dag}) \,, \nonumber \\
  {\cal H}_{os} \!& = \!& \sum_{i=1,2}\left[(\omega+\delta\omega) a_i^{\dag}a_i -\alpha E_C a_i^{\dag}a_i^{\dag}a_i a_i \right] \,, \nonumber \\
  {\cal H}_{ck} \!& = \!& -2 \alpha E_C a_1^{\dag}a_1 a_{2}^{\dag}a_{2} \,, \nonumber \\
  {\cal H}_{ch} \!& = \!& \alpha E_C \bigg( a_1a_{2}^{\dag}a_{2}^{\dag}a_{2} + a_1^{\dag}a_1^{\dag}a_1a_{2}
  \! -\! \frac{a_1^{\dag}a_1^{\dag}a_{2}a_{2}}{2} \bigg) + \! \text{H.c.} \,. \nonumber
\end{eqnarray}
In the previous expressions we introduced $\omega=1/\sqrt{\tilde{L}\tilde{C}}$, 
$E_C = e^2 /(2 \tilde{C}^2)$, $\alpha= 2L /(2L+L_J)$, and $X_{J} = [C_J/(C+2C_J)] - \alpha$.
The frequency shift $\delta \omega$ is a small correction coming from the normal 
ordering process of the nonlinearity.
The first term on the r.h.s.~of Eq.~\eqref{hamiltonian-cQED}, labeled with ${\cal H}_{lc}$,
represents an effective hopping. By choosing $C_J/(C+2C_J)= 2L /(2L+L_J)$
it is possible to make it vanish. The other terms come from the nonlinearities $\cos(\varphi_{12}/\varphi_{0})$.
In particular, ${\cal H}_{os}$ takes into account the onsite contribution (cavity frequency
and onsite Kerr terms respectively), ${\cal H}_{ck}$ describes a cross-Kerr nonlinearity,
and the term ${\cal H}_{ch}$ is a correlated hopping of photons between neighboring sites.
Each nonlinear coupling contributes with a cross-Kerr nonlinearity and an onsite Kerr nonlinearity,
where the cross-Kerr term is twice as strong as the onsite Kerr term.
For the model of two sites with one nonlinear link that we analyze here,
the cross-Kerr nonlinearity is thus twice as large as the onsite Kerr nonlinearity.
More generally, for any lattice coordination number $z$, the sum
of all cross-Kerr nonlinearities connected to a lattice site
is always twice as large as the total onsite Kerr nonlinearity on the lattice site.
This ratio can be however modified by introducing further onsite nonlinearities,
via additional superconducting qubits that locally couple to the resonators.

In our study we focus on models where interactions are short ranged,
so that only neighboring lattice sites are coupled.
To ensure that interactions decay sufficiently fast for this approximation to hold,
we require that $C_{J} \ll C$ [This approximation has been used in deriving Eq.~\eqref{hamiltonian-cQED}].
Nonetheless the cross-Kerr interaction in ${\cal H}_{ck}$, even for $X_{J}=0$
(which, for $C_{J} \ll C$, implies $\alpha \ll 1$) can be much larger
than photon losses, $2 \alpha E_C \gg \kappa$, since, {\it e.g.}, transmon qubits
have $E_{C}/h \sim 0.5$GHz and $T_{1} \sim 1\mu$s~\cite{Fedorov12}.
Note that the Josephson junctions that link two neighboring oscillators can be built
tunable by replacing them with a dc-SQUID. In this way the $E_{J}$ and thus the $L_{J}$
can be modulated by applying an external flux to the dc-SQUIDs and the Hamiltonian~\eqref{hamiltonian-cQED}
can be tuned in real time. Hence, by choosing the external flux such that $X_{J} \not=0$,
a linear tunneling of photons between the resonators can be switched on.

\begin{figure}
  \includegraphics[width = 1\linewidth]{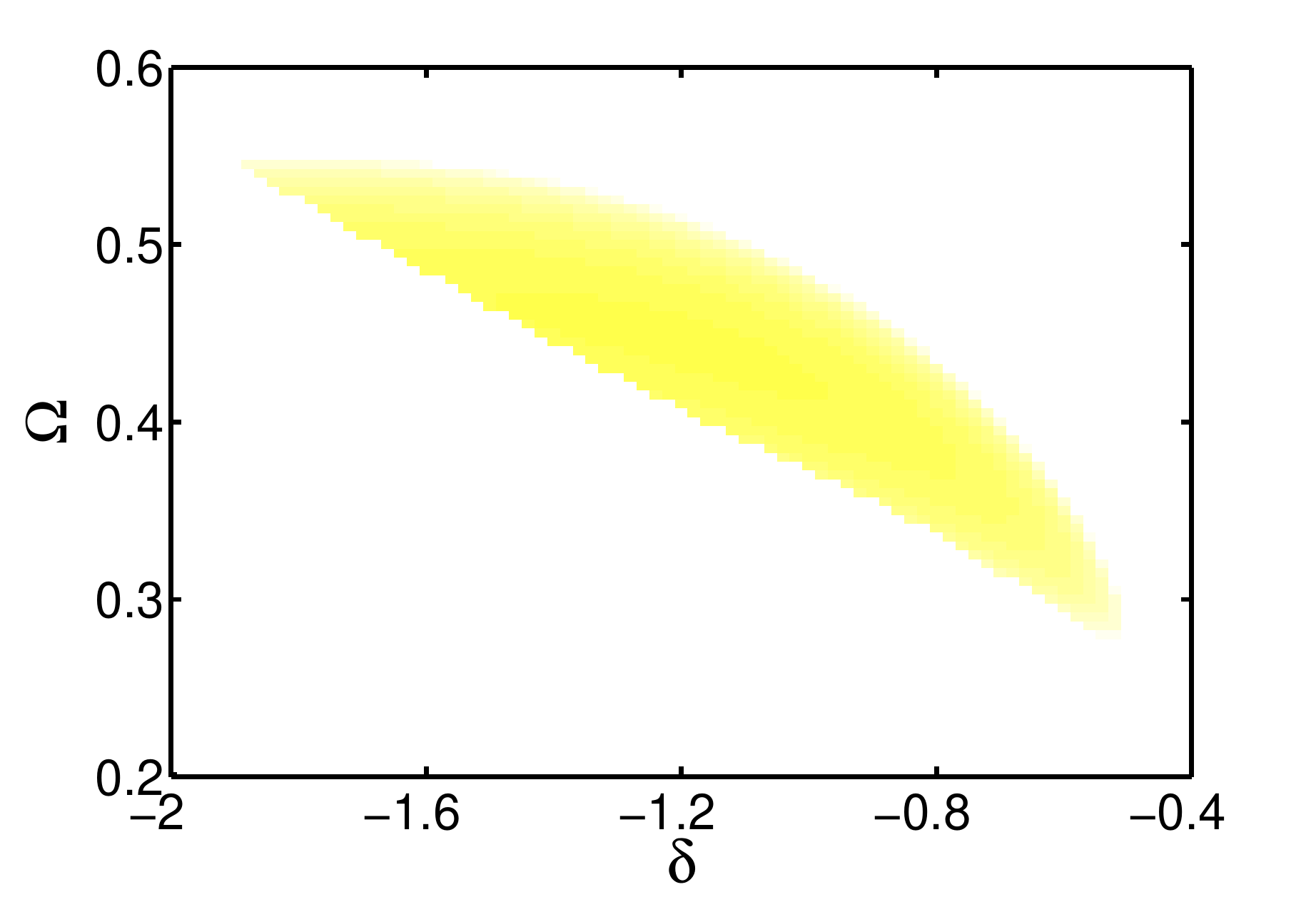}
  \caption{(Color online). Crystalline order parameter $\Delta n$ in the $\Omega$-$\delta$ plane. 
    The parameters are chosen as $zV=2U=2zJ_2=-8$, $zJ_n=4$, and $zJ=0.4$. 
    For this choice of parameters, we found $\Delta n<0.4$. 
    The ratios among these parameters correspond to those in Hamiltonian~\eqref{hamiltonian-cQED}. 
    The color code is the same as in Fig.~\ref{fig:Phase_J>0}(b).}
  \label{fig:negativeUV}
\end{figure}

The Hamiltonian derived here has the same structure as in Eq.~\eqref{modelham}. 
The main difference is that not all the coupling constants are independent. 
Moreover for this particular implementation both $U$ and $V$ are negative.
In the driven-dissipative setting we consider here, 
this however does not significantly affect the phase diagrams. 
In ground state phase diagrams, the configurations with lowest energies 
are favoured and the sign of interactions matters.
In contrast, in the driven-dissipative scenario, the drive frequency selects 
a preferred energy and interactions tend to drive the system away 
from that preferred energy, either to lower or higher energies. 
Hence configurations leading to significant interaction energies are avoided 
irrespective of the sign of the interaction. Indeed we checked that, apart
from some quantitative differences, the properties of the steady state 
phase diagram are not affected by the sign of the nonlinearities.

As already mentioned Hamiltonian~\eqref{hamiltonian-cQED}), has a fixed ratios 
of the on-site to nearest-neighbor nonlinearities and correlated hopping. 
Therefore once we fix the ratio $U/J$ the natural choice is to discuss 
the phase diagram as a function of the driving and the detuning as in Fig.~\ref{fig:Phase_U0}. 
Both parameters can be easily varied in the experiment.
The phase diagram for the circuit-QED implementation is shown in Fig.~\ref{fig:negativeUV}. 
Here we concentrate only on the transition from the uniform to the crystalline phase 
as this should the most robust feature to look at experimentally. 
The yellow region corresponds to the crystalline phase.

\section{Conclusions}
\label{conclusions}

In this work we have analyzed the phase diagram and its properties for optical
quantum many-body systems in asymptotic and stationary states,
where photon dissipation and pumping balance each other dynamically.
Besides having the practical advantage that the system in this scenario
remains stable for very long times (virtually as long as experimental conditions can be kept stable),
this is particularly interesting since such systems naturally operate out of equilibrium.
We focused on the role of cross-Kerr nonlinearities. Extending the results of Ref.~\cite{jin2013},
we analyzed in details the phase diagram in several different regimes of the coupling constants.
Furthermore we discussed the properties of the single-site density matrix in the stationary state.
In our analysis we also included the effect of correlated and pair hopping.
The most robust effect consists in the appearance of a crystalline phase
when the cross-Kerr nonlinearity becomes sizable.
Interestingly, the model can be realized even in the absence of artificial atoms inside the cavities.
Nonlinear circuits coupling neighboring resonators would suffice.
We verified that the crystalline phase survives the presence of local
quantum fluctuations by extending our analysis to a cluster mean-field.
The crystalline phase, albeit less extended, appears to be very stable.
Additional oscillating phases appear in the phase diagram.
In Ref.~\cite{jin2013} it has been suggested that in some cases this behavior
might be related to a synchronised evolution of the array.
These phases however may reveal fragile to a more accurate treatment.
The additional pair/correlated hopping slightly modify the phase boundary.
It would be interesting to explore other implementations, where
these additional couplings are more sizable, possibly leading to new phases~\cite{sowinski2012}.
We finally analyzed the implementation with circuit-QED considering specific values of the
parameters that appear for this case.

We concluded our investigation by analyzing the experimental feasibility of our proposal. 
To this aim we studied the appearance of the crystalline phase in a circuit-QED implementation. 
The phase diagram as a function of the driving and the detuning is shown in Fig.~\ref{fig:Phase_U0}.
In this respect it is important to stress that we are aware of the challenge posed 
in the realization of a cavity array.
However we would like to stress that our proposal does not introduce additional complications.
As we saw in Sec.~\ref{cQED}, the most favourable implementation is in circuit-QED.
Here, instead of putting artificial atoms inside the coplanar resonators,
one should use them to mediate the interaction between two cavities.
Although true long-range crystalline order is not possible in chains, we think that
an experiment with a one-dimensional chain of coupled cavities (minimally a ring of four cavities)
will already indicate the tendency to this type of ordering.

\acknowledgments

We acknowledge fruitful discussions with A. Tomadin.
This work was supported by EU through IP-SIQS
by DFG through the Emmy Noether project HA 5593/1-1 and the CRC 631,
by Italian MIUR via PRIN Project 2010LLKJBX and FIRB Project RBFR12NLNA,
and by National Natural Science Foundation of China under Grant No. 11175033
and No. 11305021.

\end{document}